\documentclass{article}
\usepackage{xspace}
\usepackage{xcolor}
\usepackage{float}
\usepackage{amsmath, amssymb}
\usepackage{gensymb} 
\usepackage{bm}
\usepackage{tikz}

\newcommand{\co}{CO$_2$\xspace}

\author{Odd Andersen$^\ast$ \and Halvor M{\o}ll Nilsen \and Sarah Gasda}
\title{Modelling Geomechanical Impact of CO$_2$ Injection Using Precomputed Response Functions}

\begin{document}
\maketitle
 \begin{abstract}
   When injecting \co or other fluids into a geological formation, pressure
   plays an important role both as a driver of flow and as a risk factor for
   mechanical integrity.

   The full effect of geomechanics on aquifer flow can only be captured using a
   coupled flow-geomechanics model.  In order to solve this computationally
   expensive system, various strategies have been put forwards over the years,
   with some of the best current methods based on sequential splitting.

   In the present work, we seek to approximate the full geomechanical effect on
   flow without the need of coupling with a geomechanics solver during
   simulation, and at a computational cost comparable to that of an uncoupled
   model.  We do this by means of precomputed pressure response functions.  At
   grid model generation time, a geomechanics solver is used to compute the
   mechanical response of the aquifer for a set of pressure fields.  The
   relevant information from these responses is then stored in a compact form
   and embedded with the grid model.

   We test the accuracy and computational performance of our approach on a
   simple 2D and a more complex 3D model, and compare the results with those
   produced by a fully coupled approach as well as from a simple decoupled
   method based on Geertsma's uniaxial expansion coefficient.
 \end{abstract}
 \clearpage

\newcommand{\rev}[1]{\textcolor{red}{#1}}
\newcommand\numberthis{\addtocounter{equation}{1}\tag{\theequation}}

\newcommand{\dt}{\frac{\partial}{\partial t}}
\newcommand{\stot}{\boldsymbol\sigma}
\newcommand{\I}{\textbf{I}}
\newcommand{\beps}{\boldsymbol\epsilon}
\newcommand{\bu}{\textbf{u}}
\newcommand{\bq}{\textbf{q}}
\newcommand{\bg}{\textbf{g}}
\newcommand{\bt}{\textbf{t}}
\newcommand{\bG}{\textbf{G}}
\newcommand{\bS}{\textbf{S}}
\newcommand{\bF}{\textbf{F}}
\newcommand{\bP}{\textbf{P}}
\newcommand{\bQ}{\textbf{Q}}
\newcommand{\bM}{\textbf{M}}
\newcommand{\bE}{\textbf{E}}

\newcommand{\blockmatrix}[3]{
\begin{minipage}[t][#2][c]{#1}%
\center%
#3%
\end{minipage}%
}%
\newcommand{\fblockmatrix}[3]{%
\fbox{%
\begin{minipage}[t][#2][c]{#1}%
\center%
#3%
\end{minipage}%
}%
}

\section{Introduction}

\subsection{Background}

In order for carbon capture and storage (CCS) to play a significant role in the
mitigation of climate change, hundreds or even thousands of megatonnes of would
have to be injected annually into geological formations on a worldwide basis
\cite{CBNB15:wrr}.  This represents a huge scale-up from current practice and
experience, which means that we to a large extent need to rely on theoretical
knowledge and computer simulations to gain insight into storage-related issues
such as injectivity, capacity or long-term migration. Frequently, the use of
computer models to investigate such questions will involve working with sparse
or poorly constrained data, and address issues that relate to a wide range of
spatial and temporal scales.  The ability to run a large number of simulations
in a reasonable amount of time is crucial, as it allows efficient exploration of
different hypotheses and choices of parameters, evaluation of various injection
scenarios and assessment of potential risk factors.  Examples are workflows that
include inverse modeling, optimization of storage operations, or risk analysis
in the face of a large number of unknown or uncertain parameters.  For this
reason, the case for simplified tools for modeling \co storage, less demanding
than the use of traditional reservoir simulators, has previously been argued
\cite{NordbottenCelia, co2lab:part4}.  Significant effort has been spent on the
development of simplified tools for predicting subsurface flow of \co at large
scales.  Over time, the capabilities of such tools have been expanded to account
for a considerable range of physical phenomena \cite{ND11:wrr, WRCR:WRCR12618,
  Gasda2013, co2lab:part3, Andersen:14}.  However, reduced models that properly
account for the two-way coupling between fluid flow and rock mechanics have so
far received less attention, although recent work in this field includes
the linear vertical deflection model of \cite{WRCR:WRCR21923}.

Geomechanical issues have proved to be important even in the context of current,
limited size \co storage operations \cite{verdon2013comparison, eiken:11}, and
will be even more important to understand for the larger-scale operations
envisioned in a global mitigation scenario.  Potential geomechanical risks
associated with \co storage include seismicity, fault reactivation, rock
fracturing and unwanted fluid displacement \cite{rutqvist2012}.  Understanding
these risks will be important for any discussion regarding injectivity, storage
capacity or long-term safety.

Investigating geomechanical effects requires the ability to model the interplay
between fluid flow, pressure, mechanical stresses and strains, as well as the
impact of deformation on rock properties.  The theoretical framework of
poromechanics was first introduced in the forties by the work of Karl von
Terzaghi and Maurice Biot.  Significant attention has later been paid to how
this framework can be applied to reservoir engineering, and how the resulting
system of coupled equations describing pressure and mechanical displacement can
be efficiently solved \cite{settari1998,
  settari2001,longuemare2002geomechanics, dean2006comparison, kim2010sequential,
  mikelic2013convergence, girault2015convergence}.

A significant number of studies have been carried out on the topic of
geomechanics and \co injection over the years.  At the theoretical level, the
poromechanical framework has been used to investigate the risk of fracturing or
fault slip caused by elevated pressure and reduced effective stress
\cite{Rutqvist2002,rutqvist2007estimating, cappa2012seismic}, and assess the
risk of induced seismicity \cite{GRL:GRL28363}.  A reduced model for coupled
flow and geomechanics, adapted to the case of \co storage, was proposed by
\cite{WRCR:WRCR21923}, whereas \cite{Darcis:13} situates geomechanics within a
comprehensive modeling framework for \co storage, where different physical
processes are modeled depending on the spatial and temporal scale.  Specific
case-studies involving \co injection and geomechanical impacts include Ketzin
\cite{ouellet2011reservoir}, In-Salah \cite{rutqvist2010coupled,
  preisig2011coupled, bissell2011full}, Vedsted \cite{tillner2014coupled} and
the CO2CRC Otway Project \cite{aruffo2014geomechanical}.

When fluid is injected into a geological formation, the resulting change in
underground pore pressure is accompanied by some degree of mechanical
deformation of the rock matrix caused by a change in the underground balance
between mechanical and pressure forces.  The rock deformation
also has an impact on the evolution of the pressure field itself, as the
expansion or contraction of the rock matrix modifies material parameters that
affect fluid flow, in particular porosity and permeability
\cite{davies1999stress}.  Whereas the influence of pressure on rock deformation
is fundamental in any combined simulation of geomechanics and fluid flow, the
impact of geomechanics on the pressure field is often simplified or neglected.
In reservoir modeling, the most common practice is to simulate reservoir flow in
isolation, and compute the corresponding mechanical deformation as a
post-processing step if desired.  Under this approach, the full impact of rock
mechanics on fluid flow is not explicitly modeled, but the effect is approximated
by making rock properties (typically linear) functions of local pressure or
assumed stress.  In particular, the effect of rock expansion is frequently
modeled using a pore volume compressibility coefficient, which affects the
accumulation term of the governing mass-balance equations \cite{settari1998,
  dean2006comparison, longuemare2002geomechanics}.  This approach is standard
practice in most commercial and academic reservoir simulation software, and is
often considered to provide a perfectly adequate approximation of the real
behavior of the poromechanical system.  In the context of geomechanics and \co
storage, it has been utilized in e.g.  \cite{ouellet2011reservoir,
  aruffo2014geomechanical, tillner2014coupled}.

On the other hand, a full account of geomechanical effects requires explicit
modeling of the two-way coupling between the flow and mechanical
subsystems. This requires embedding the reservoir simulation grid within a
larger mechanical grid that includes the over- and underburden, and solving the
equations of the combined system together.  In this setting, flow is frequently
restricted to the reservoir part of the model, whereas mechanical deformations
occur throughout the model.  

Under the fully coupled approach, each discrete element (node, face or cell,
depending on the numerical discretization) in the mechanical model comes with
three displacement unknowns that must be solved for, alongside with the unknowns
of the flow equation.  The full flow/mechanical system is thus described by two
coupled equations of elliptic character.  All in all, this leads to a numerical
model with a large number of unknowns, which is computationally heavy
to solve.  Various strategies based on sequential splitting have been proposed
for computing the solution in an efficient manner, e.g.
\cite{settari2001,longuemare2002geomechanics, dean2006comparison}, and the
``fixed-stress split'' approach has been shown to have particularly favorable
convergence properties \cite{kim2010sequential, mikelic2013convergence}.  In
addition to being attractive from a computational viewpoint, sequential
splitting approaches have the additional advantage that they allow flow and
mechanics equations to be separately addressed by standard solvers that have
been highly adapted to their specific fields over the years.  Examples where
fully coupled geomechanical modeling has been used in the context of \co storage
include \cite{Rutqvist2002}, \cite{Darcis:13} and \cite{preisig2011coupled}.

The general need for fully coupled models versus the more common practice of
one-way coupling has been argued both ways in the past in the context of aquifer
pumping and subsidence problems \cite{lewis1991coupling, NAG:NAG1610161105}.
Regarding \co injection, the geomechanical feedback on fluid flow was observed
to be weak and localized around the injection well for the test cases explored
in \cite{Darcis:13}.  In any case, there seems to be a general recognition in
the research community that the full geomechanical impact on flow is of
relevance in some situations, as brought to witness by the significant amount of
related research activity in recent years.

\subsection{Our proposed approach}

Sequential splitting strategies in combination with iterative linear solvers can
provide a significant gain in computational efficiency compared to solving all
equations as a single system.  However, they remain significantly more expensive
than modeling aquifer flow in isolation, as is common practice in
reservoir engineering.  This is because the sequential splitting approach still
requires the mechanics equations to be solved repeatedly throughout the
simulation.  In the case of iterative strategies, both the flow and mechanics
equations have to be solved multiple times for every timestep until
convergence is achieved.  Moreover, even though the splitting approach permits
the flow simulator and geomechanics solver to be chosen independently, it is
still necessary to have access to both types of software at runtime.

In the approach we propose in this paper, we aim to include the full effect of
geomechanics on fluid flow to different degrees of approximation, without the
need of a mechanics solver at simulation time.  As such, the computational
requirement of a flow simulation becomes comparable to that of standard
reservoir modeling, but can still reproduce the solution obtained from solving
the two-way coupled system.  This advantage comes at the expense of a
precomputation step at grid generation time, where the mechanical responses of
the grid for a large set of discrete pressure fields are determined and stored
for re-use at flow simulation time.  The approach is based on the theory of
linear poroelasticity \cite{wang2000theory}.  Its practical feasibility relies
on the observation that although a local change in a pressure field can have
significant non-local mechanical effects in terms of \emph{displacements}, the
impact on rock expansion (i.e. the \emph{divergence} of displacements), is
typically much more local in the case of boundary conditions relevant to \co
injection, as will be discussed in the next section.

For simplicity, our focus is limited to one-phase flow, although the proposed
method is trivially extendable to multi-phase flow under the assumption that
mechanical response is tied to \emph{effective} fluid pressure.  Also, we have omitted
the impact of geomechanical stress on rock permeability.  This effect could also
be easily included in the model, but merits a separate analysis and is thus
considered future work.

\section{Conceptual system, theory and basic idea}

We consider a fluid injection or extraction operation into/out from a
subterranean reservoir or aquifer.  We neglect fluid flow through aquifer top
and bottom boundaries, and impose either no-flow or fixed pressure along lateral
boundaries.  To model mechanical deformation, the aquifer model is embedded in a
larger rock matrix that includes the over- and underburden (collectively
referred to as the \emph{surrounding domain}).  The overburden zone extends up to
the surface, where the mechanical boundary condition is that of fixed, normal
traction, while the underburden zone extends down to some specified deeper
level where zero mechanical displacements are assumed.  The lateral boundary
conditions can be of any type; for the purpose of the examples presented in this
paper, we use the `roller' type with zero lateral displacements and constant
vertical stress.

\begin{figure}[!htb]
  \centering
  \includegraphics[width=0.35\columnwidth, clip=true, trim=90 50 70 30]{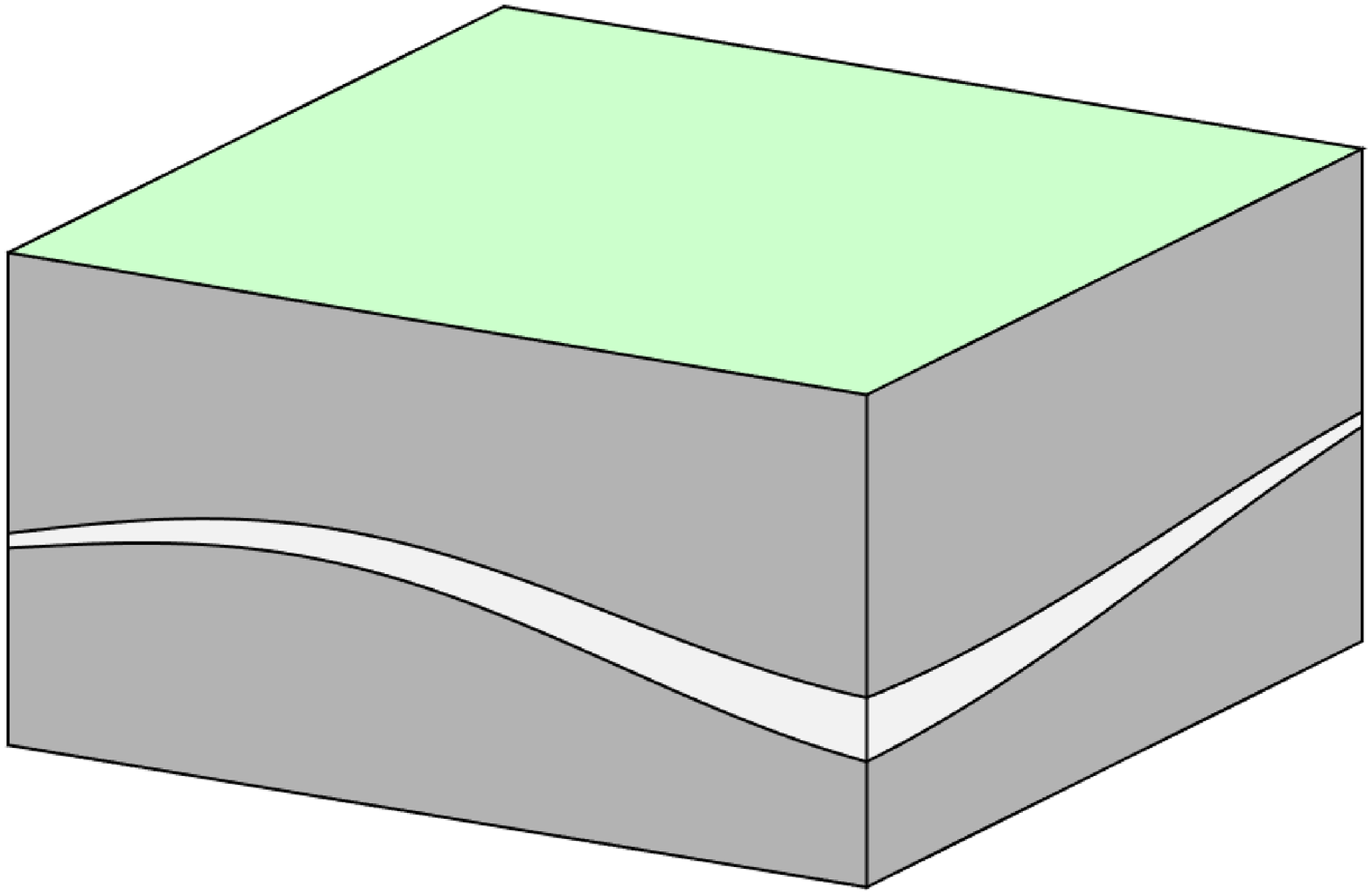}
  \includegraphics[width=0.35\columnwidth, clip=true, trim=40 90 80
  30]{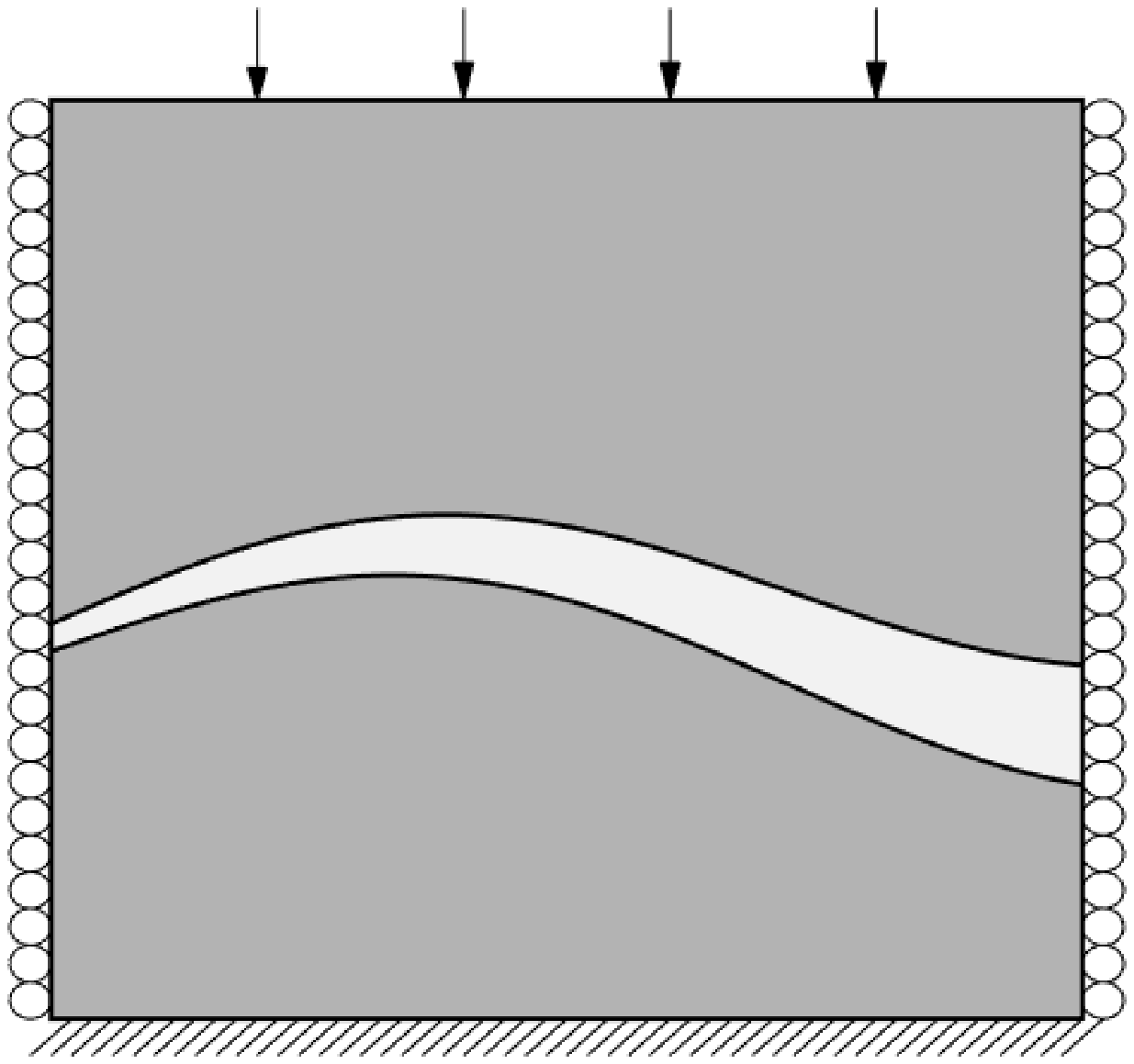}
  \caption{\emph{Left:} The conceptual model.  The aquifer is illustrated as a
    light band within a darker matrix of surrounding rock that extends up to
    the ground level (green surface).  Fluid flow only takes place in the
    aquifer, whereas mechanical deformations are computed for the full model.
    \emph{Right:} Cross-section view of the model, with mechanical boundary
    conditions indicated.  Lateral boundaries can be of any type (`roller' type
    indicated on figure).}\label{fig:model}
\end{figure}

The complete model domain is denoted $\Omega$, with boundary $\partial\Omega$.
The domain $\Omega$ is subdivided in two separate zones, the aquifer
$\Omega_{aq}$ and the surrounding domain $\Omega_{sd}$.  For simplicity, we
model $\Omega_{aq}$ and $\Omega_{sd}$ as having separate but spatially invariant
sets of poroelastic moduli.  As the modeling of fluid flow is restricted to the
aquifer, the associated pressure field is only defined on $\Omega_{aq}$.
Accordingly, drained poroelastic moduli are used to describe
$\Omega_{aq}$, whereas $\Omega_{sb}$ is described in terms of undrained
moduli.  We thereby neglect the effect of pressure changes and flow occurring
\emph{outside} the aquifer as a result of mechanical deformations induced by the
pressure field \emph{inside} the aquifer.  A basic, layered structure similar to
Figure~\ref{fig:model} is assumed, but  $\Omega_{aq}$ and $\Omega_{sd}$ can be
of arbitrary geometrical shape and topology, including pinch-outs.

\subsection{Linear poroelasticity formulation}

Since mechanical deformations are expected to remain small and mechanical
equilibrium assumed to be reached on a time scale much shorter than that of
fluid flow, we model the coupled flow-mechanical system within the framework of
linear poroelasticity.  The governing equations then consist of the force
equilibrium equations for mechanics and an inhomogeneous diffusion equation for
fluid pressure.  Coupling between equations arise as the gradient of pore
pressure plays the role as a body force in the mechanical equilibrium equations,
whereas mechanical strain appears in the accumulation term of the pressure
diffusion equation.  In our formulation, the independent unknown variables
consist of the mechanical displacement field $\textbf{u}=[u_x, u_y, u_z]^T$
defined on $\Omega$, and the fluid pressure $p$ defined on $\Omega_{aq}$.

The mechanical system is assumed to be in translational and rotational
equilibrium at any time.  Rotational equilibrium implies that the total stress
tensor, $\stot$, should be symmetrical, i.e.:
\begin{equation}\label{eq:symeq}
  \boldsymbol{\sigma} = \boldsymbol{\sigma}^T
\end{equation}
whereas translational equilibrium requires that the divergence of this tensor
counterbalances the body forces $\textbf{F}$ acting on it:
\begin{equation}\label{eq:transeq}
  \nabla\cdot\stot + \textbf{F} = 0
\end{equation}
In linear poroelasticity, provided tensile stresses are taken to be positive,
total stress equals the difference between \emph{effective stress} $\stot'$ and
a term proportional to fluid pressure:
\begin{equation}\label{eq:stot-def}
  \stot = \stot' - \alpha p\I
\end{equation}
where $\alpha$ is the \emph{Biot-Willis coefficient} and $\I$ is the identity
matrix in $\mathbb{R}^3$.  Effective stress is linked to elastic strain
$\beps$ through Hooke's law: 
\begin{equation}\label{eq:hooke}
  \stot' = \textbf{C}\beps
\end{equation}
where $\textbf{C}$ is the fourth-order elasticity tensor, and the elastic strain
tensor $\beps$ is defined as:
\begin{equation}\label{eq:strain}
  \beps = \frac{1}{2}(\nabla\bu + \nabla\bu^T)
\end{equation}
For an isotropic material, \eqref{eq:hooke} reduces to:
\begin{equation} \label{eq:isotropic-hooke}
  \stot' = 2G\beps + (K - \frac{2}{3}G)\text{tr}({\beps})\I
\end{equation}
where $K$ and $G$ denote respectively the drained bulk and shear moduli of the
material, and $\text{tr}(\beps)$ denotes the trace of $\beps$. $K$ and $G$ can
be space-dependent.  By combining the equilibrium equation \eqref{eq:transeq}
with \eqref{eq:stot-def}, \eqref{eq:strain} and \eqref{eq:isotropic-hooke}, we
obtain the displacement formulation of the force balance equation for an
isotropic material:
 \begin{equation}\label{eq:forcebalance}
   \nabla\cdot (G\nabla\textbf{u}) + \nabla\left((K +
   \frac{1}{3}G)\nabla\cdot\textbf{u}\right) - \alpha\nabla p = \rho_b\textbf{g}
 \end{equation}
Here, we have considered that the body force $\textbf{F}$ from
\eqref{eq:transeq} consists of the gravity force only, i.e. $\textbf{F} =
-\rho_b\textbf{g}$, where $\rho_b$ is the bulk density of the medium and
$\textbf{g}$ is gravitational acceleration.  

In addition, boundary conditions must be specified.  Boundary conditions can be
of different type for different spatial components of the displacement vector
$\bu$.  For instance, lateral roller boundary conditions specify
zero displacement (Dirichlet) for $u_x$ and $u_y$, but fixed-stress (Neumann)
for $u_z$.  Each spatial component $i\in{\{x, y, z\}}$ of $\bu$ thus has its own
subdivision of $\partial\Omega$ into a Neumann ($\Gamma_t^i$) and a Dirichlet
($\Gamma_g^i$) part.  For each spatial component $i$, fulfillment of the boundary
conditions requires:
 \begin{align}\label{eq:mech-bcond}
   u_i &= g_i \text{ on } \Gamma^i_g\\
   \sigma_{ji}n_j &= t_i \text{ on } \Gamma^i_t
 \end{align}
 Regarding rotational equilibrium, from \eqref{eq:stot-def}, \eqref{eq:strain}
 and \eqref{eq:isotropic-hooke} it is easy to see that the symmetry requirement
 of \eqref{eq:symeq} is automatically fulfilled.

The pressure equation governing single-phase fluid flow is obtained by combining
the fluid continuity equation with Darcy's constitutive relationship between
pressure and flow.  The fluid continuity equation with volumetric source term $Q$ can be written:
\begin{equation}\label{eq:continuity}
\dot\zeta + \nabla\cdot\bq = Q
\end{equation}
Here, $\bq$ represents the volumetric fluid flux and $\zeta$ denotes the
accumulation term.  In poroelastic literature, starting with \cite{biot:41},
$\zeta$ is commonly referred to as the \emph{increment of fluid content}, and
represents the volume of fluid imported into a control volume, per control
volume.  It is modeled as depending linearly on fluid pressure $p$ and volumetric
strain $\epsilon = \text{tr}(\beps) = \nabla\cdot\bu$, so that the time
derivative becomes:
\begin{equation}\label{eq:zetader}
  \dot\zeta = \frac{\partial\zeta}{\partial p}\dot p +
  \frac{\partial\zeta}{\partial\epsilon}\dot\epsilon = S_\epsilon\dot p + \alpha\dot\epsilon
\end{equation}
Here $S_\epsilon$ is called the \emph{specific storage coefficient at constant
  strain} and $\alpha$ is the Biot-Willis coefficient that was already introduced
in \eqref{eq:stot-def}. The following expression can be derived for $S_\epsilon$
\cite{wang2000theory}:
\begin{equation}\label{eq:s-epsilon}
  S_\epsilon = \frac{1}{K}(1-\alpha)(\alpha-\phi) + \frac{\phi}{K_f}
\end{equation}
Here, $\phi$ is the porosity of the medium and $\frac{1}{K_f}$ represents fluid
compressibility.

The flux $\bq$ is linked to the fluid pressure through Darcy's law:
\begin{equation}
  \bq = -\frac{k}{\mu}(\nabla p - \rho_f\bg)
\end{equation}
where $k$ is the permeability of the medium, $\mu$ represents fluid viscosity
and $\rho_f$ fluid density.  Combining Darcy's law with \eqref{eq:continuity} and
\eqref{eq:zetader}, we obtain the pressure equation for single-phase flow:
\begin{equation}\label{eq:pressure-equation}
  \alpha\dot\epsilon + S_\epsilon\dot p - \nabla\cdot\frac{k}{\mu}(\nabla p - \rho_f\bg) = Q
\end{equation}

In our model, this equation governs fluid flow in $\Omega_{aq}$.  Fluid
flow is neglected in $\Omega_{sd}$, where we instead use the undrained bulk modulus $K_u =
K + \frac{\alpha^2}{S_\epsilon}$, and the corresponding force
balance equation reduces to that of linear elasticity.

The combined poroelastic equation system, with $\bu$ and $p$ as unknowns, becomes:
\begin{align}
    &\nabla\cdot(G\nabla\textbf{u}) + \nabla\left((K + \frac{1}{3}G)\nabla\cdot\textbf{u}\right) -
    \alpha\nabla p = \rho_b\textbf{g} \;\;\;&\text{ in } \Omega_{aq} \label{eq:sys1}\\
   &\nabla\cdot(G\nabla\textbf{u}) + \nabla\left((K_u + \frac{1}{3}G)\nabla\cdot\textbf{u}\right) =
   \rho_b\textbf{g} &\text{ in } \Omega_{sd}\label{eq:sys2}\\
   &\alpha\dot\epsilon + S_\epsilon\dot p - \nabla\cdot\frac{k}{\mu}(\nabla p - \rho_f\bg) =
   Q &\text{ in } \Omega_{aq}\label{eq:sys3}
\end{align}
Poroelastic parameters are here allowed to be spatially heterogeneous.  In
particular, they may vary between the aquifer and its surroundings, or between
different geological layers.  We see that \eqref{eq:sys1} is
coupled to \eqref{eq:sys3} through the term $\alpha\nabla p$, whereas
\eqref{eq:sys3} is coupled to \eqref{eq:sys1} and \eqref{eq:sys2} through the
term $\alpha\dot\epsilon$ $(= \alpha\nabla\cdot\dot\bu)$.

Mechanical boundary conditions at $\partial\Omega$ are given by
\eqref{eq:mech-bcond}. Material continuity dictates the boundary conditions at
the interfaces between $\Omega_{aq}$ and $\Omega_{sd}$.  Boundary conditions for
flow in $\Omega_{aq}$ are:
\begin{align}\label{eq:flow-bcond}
  p &= p^0 \text{ on } \Gamma_p\;\;\;&\text{(constant pressure)}\\
  \bq &= 0 \text{ on } \Gamma_{\bq}\;\;\;&\text{(no-flow)}
\end{align}
where it is understood that the top and bottom boundaries of the aquifer are
both part of $\Gamma_{\bq}$.

\subsection{timestepping and linear system}

A backwards-Euler time discretization of \eqref{eq:sys3} yields:
\begin{equation}
  \alpha \nabla\cdot\boldsymbol{u^{n+1}} + S_\epsilon p^{n+1} - \Delta
    t\frac{K}{\mu}\nabla^2 p^{n+1} = \Delta t Q^{n+1} +
    \alpha\nabla\cdot\boldsymbol{u^n} + S_\epsilon p^n 
\end{equation}
where $\Delta t$ is the timestep size, and $(\bu^{n}, p^{n})$ and $(\bu^{n+1},
p^{n+1})$ are the values of $(\bu, p)$ at timestep $n$ and $n+1$ respectively.
Given an applicable spatial discretization scheme, the resulting linear system
can be expressed as:
\begin{equation}
  \begin{bmatrix} \bG & \bS + \Delta t \bP \end{bmatrix}
  \begin{bmatrix}\bu^{n+1}\\p^{n+1}\end{bmatrix} =
  \begin{bmatrix} \bG & \bS\end{bmatrix}
  \begin{bmatrix}\bu^{n}\\p^{n}\end{bmatrix} + \Delta t \bQ
\end{equation}
where $\bG$ is a discretization of $\alpha\nabla\cdot\bu$ restricted to aquifer
cells, $\bS$ is a discretization of $S_\epsilon p$, $\bP$ is a discretization of
$\frac{K}{\mu}\nabla^2p$ and $\bQ$ is a discretization of the source term $Q$.
If we moreover use $\bM$ to represent a discretization of $\nabla\cdot(G\nabla\textbf{u}) +
\nabla\left((K + \frac{1}{3}G)\nabla\cdot\textbf{u}\right)$ and $\bF_u$ to represent a
combined discretization of body forces $\rho_b\bg$ and boundary forces $\bt$, we
get the linear system for the complete poroelastic problem:
\begin{equation}\label{eq:discretized}
  \begin{bmatrix} \bM & \bG^T \\\bG & \bS + \Delta t \bP \end{bmatrix}
  \begin{bmatrix}\bu^{n+1}\\p^{n+1}\end{bmatrix} = 
  \begin{bmatrix} \textbf{0} & \textbf{0}\\\bG & \bS\end{bmatrix}
  \begin{bmatrix}\bu^{n}\\p^{n}\end{bmatrix} +
  \begin{bmatrix}\bF_u\\ \Delta t \bQ \end{bmatrix} 
\end{equation}
Since the gradient operator is the negative adjoint of the divergence operator
away from the boundary, the discretization of $\alpha \nabla p$ is here $-G^T$.
It is assumed that the matrices $\bM$, $\bS$ and $\bP$ are symmetrical. 

It is worth noting that the number of aquifer cells in a discrete model,
$N_{aq}$, is usually significantly less than the total number of cells in the
model $N = N_{aq} + N_{sd}$.  Moreover, since pressure is scalar whereas $\bu$
has 3 components (in 3D), the number of discrete values $u_i$ is approximately
$3N$, whereas the number of discrete values $p_i$ is only about $N_{aq}$.  As a
consequence, the square matrix $\bM$ is much larger than $\bS+\Delta t\bP$ and
$\bG$, and the system block matrix takes on the following shape: \\
\begin{center}
\begin{tabular}{llll}
\fblockmatrix{1.5in}{1.5in}{\bM}&
\fblockmatrix{0.4in}{1.5in}{$\bG^T$}&
\\
\fblockmatrix{1.5in}{0.4in}{\bG} & 
\fblockmatrix{0.4in}{0.4in}{$\bS+\Delta t \bP$}
\\
\end{tabular}
\vspace{0.5cm}
\end{center}

Another important observation is that the influence of the displacement field
$\bu$ on the pressure equation is only in terms of its divergence $\epsilon$.
As we argue in the next subsection, the impact of a local pressure change on
$\epsilon$ is generally localized, even though the influence on $\bu$ can be
far-reaching.  This is key to the practicality of our proposed method when
applied to large models.

Finally, we note that if one were to eliminate $\bu$ from \eqref{eq:discretized}
using the Schur complement, we define {$\bE = -\bG\bM^{-1}\bG^T$} and obtain:

\begin{equation}\label{eq:schur}
  (\bE + \bS + \Delta t\bP)p^{n+1} = (\bE + \bS)p^n + \Delta t \bQ
\end{equation}


The matrix $\bE\in\mathbb{M}_{N_{aq}}$ here represents volumetric strain
$\epsilon$ as a function of $p$ only.  It is a symmetric matrix whose size is
significantly smaller than the full system matrix ($\mathbb{M}_{3N+N_{aq}}$).
On the other hand, $\bE$ is non-sparse, making it impractical to store and
invert directly.  As becomes apparent in the discussion of our method below, our
approach can be numerically interpreted as the use of a truncated form of the
Schur complement.
 
\subsection{Decoupled flow simulation and Geertsma's uniaxial poroelastic
  expansion coefficient}

Under the assumption of zero lateral strain and constant vertical stress, the
pressure equation \eqref{eq:sys3} completely uncouples from the mechanical
equations \eqref{eq:sys1} and \eqref{eq:sys2} \cite{wang2000theory}. With this
assumption, which is often made in hydrogeology, changes in local strain $
\Delta\epsilon$ become directly proportional to changes in local pressure
$\Delta p$ through the relation:
\begin{equation}\label{eq:uniaxial}
  \Delta \epsilon = c_m \Delta p
\end{equation}
The factor $c_m$ is called
\emph{Geertsma's uniaxial poroelastic expansion coefficient}.  It is related to
the elastic moduli $K$ and $G$ as follows: 
\begin{equation}
  c_m = \frac{\alpha}{K_v}
\end{equation}
where $K_v = K + \frac{4}{3}G$ is the (drained) \emph{uniaxial bulk modulus} of
the material.   By combining \eqref{eq:uniaxial} with \eqref{eq:sys3}, we obtain
the uncoupled pressure equation:
\begin{equation}\label{eq:local}
  (\alpha c_m + S_\epsilon)\dot p - \frac{K}{\mu}\nabla^2p=Q
\end{equation}
The only unknown in this equation is pressure.  This is equivalent to the
standard transient flow equation used in hydrogeology:
\begin{equation}
  S\dot p -\frac{K}{\mu}\nabla^2p=Q\label{eq:uncoupled-flow}
\end{equation}
where $S = \alpha c_m + S_\epsilon$ is the \emph{uniaxial specific storage
  parameter}.  Combining \eqref{eq:local} with the expression for $S_\epsilon$
in \eqref{eq:s-epsilon}, we see that $S$ can be divided into two parts $S = S_1
+ S_2$, where $S_1 = \alpha c_m + \frac{1}{K}(1-\alpha)(\alpha-\phi)$ accounts
for rock expansion and grain compressibility, whereas $S_2 = \frac{\phi}{K_f}$
accounts for fluid compressibility.  Such a presentation of the accumulation term is
frequently made in commercial and academic reservoir simulation code (although
not necessarily derived in the same way), where the equivalent of 
$S_1$ is interpreted as a pore volume compressibility coefficient, and $S_2$ can
be easily generalized to the nonlinear case where fluid density is obtained from
an equation of state.

We refer to this decoupled approach, popular in reservoir modeling, as the
\emph{local model}, since the relationship between aquifer volumetric strain and
pressure \eqref{eq:uniaxial} is completely local. After solving for pressure
using \eqref{eq:local}, mechanical displacements can be subsequently obtained
using \eqref{eq:sys1} and \eqref{eq:sys2}, in which case it can be considered a
one-way coupled approach.

As demonstrated in the next section, the local model often produces good results
despite its simplicity and low computational requirements.  Nevertheless, it is
clear that the assumption of zero lateral strain can be only approximately true
throughout a problem domain.  The assumption breaks down in the presence of
strong lateral pressure gradients, e.g. in the vicinity of injecting or
producing wells.  The model's tendency to produce good results even in the
presence of strong pressure gradients can also be understood through a different
interpretation.  As shown in the Appendix, the analytical solution of
\eqref{eq:forcebalance} in an infinite domain with constant elastic moduli leads
to relation \eqref{eq:uniaxial} without any further assumptions on stresses and
strains.  In other words, any changes in strain that are non-locally dependent
on pressure must in some way be related to material heterogeneities and/or
influence from the imposed boundary conditions.  (For instance, if the system as
a whole is not allowed to expand, internal changes in volume would always be
zero-sum, so a local expansion of a given internal volume would necessarily
have non-local impact).  In any case, as the system tends towards steady state,
the time-dependent term in \eqref{eq:sys3} vanishes and the flow and mechanical
systems decouple, in which case \eqref{eq:sys3} and \eqref{eq:uncoupled-flow}
become equivalent.

\begin{figure}[!h]
  \centering
  \includegraphics[width=0.7\columnwidth]{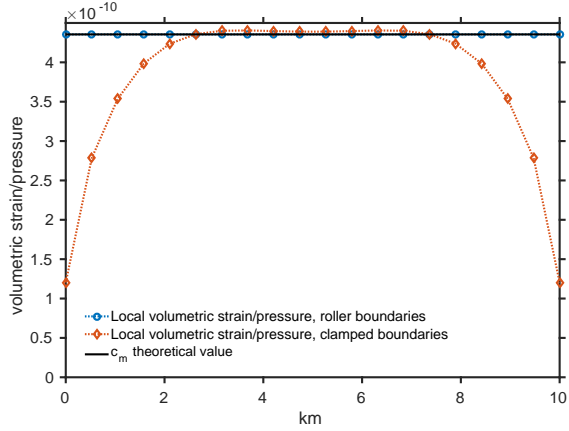}
  \caption{Numerical estimation of Geertsma's uniaxial poroelastic expansion
    coefficient $c_m$ for a two-dimensional (x, z) flat and horizontal aquifer.
    The x-axis represents spatial position along the aquifer, whereas the y-axis
    represents the local change in volumetric strain for a unit, global pressure
    increase throughout the aquifer domain.}\label{fig:geertsma}
\end{figure}

Since $c_m$ is derived from the assumption of zero lateral strain, it represents
the increase in pore volume resulting from a uniform vertical expansion of the
aquifer, as would approximately result from a uniform pressure increase
throughout the aquifer domain. This means that we can use our poroelastic
formulation to compute a numerical generalization of $c_m$, which we note $\bar
c_m$, and which can account for arbitrary boundary conditions and heterogeneous
rock parameters.  This is done by solving the elastic equations \eqref{eq:sys1},
and \eqref{eq:sys2} a single time for $p$ equal to a uniform, unit pressure
increase, and using the obtained displacement field to compute the corresponding
volumetric strain $\epsilon$ for each individual aquifer cell.  An example is
presented in Figure~\ref{fig:geertsma}, where we have computed $\bar c_m$ using
a 2D (x, z) model of a horizontal aquifer embedded in a larger rock matrix, at a
depth of 950 m.  The aquifer is 10 km long and 100 m thick, and the poroelastic
parameters are $\alpha=0.9$, $K_{aq}=10^9$, $G_{aq}=8\times 10^8$,
$K_{sd}=1.3\times 10^{11}$ and $G_{sd}=1.4\times 10^{10}$.  We plot
$\Delta\epsilon/\Delta p$ for two different specified lateral boundary
conditions, and compare with the theoretical value of $c_m$.  With roller
boundary conditions, $\bar c_m$ here reproduces the theoretical value of $c_m$,
whereas clamped boundaries leads to strong variation of $\bar c_m$ towards the
lateral boundaries, reflecting the local breakdown of the assumption of zero
lateral strain and constant vertical stress.  The numerically obtained parameter
$\bar c_m$ is thus able to model more general boundary conditions than the underlying
assumptions behind the analytical parameter $c_m$ suggests.  As an extreme
example, $\bar c_m$ would equal zero for a materially uniform domain constrained with
zero-displacements on all sides.  Material heterogeneities are also 
taken into account in the numerical value of $\bar c_m$. 

\subsection{The use of precomputed response functions}

The local model just described relies on simplifying assumptions that are at
best approximately true, and fails to capture non-local relations between
pressure and volumetric strain.  On the other hand, solving the fully coupled
linear poroelastic system represented by \eqref{eq:discretized} quickly becomes
computationally demanding.  We here seek to establish a method that retains much
of the computational efficiency of the local model, that does not require any
assumptions on strains or stresses, and that is capable of producing results
that are close to those obtained by a fully-coupled model.

The force balance equation of linear poroelasticity \eqref{eq:forcebalance}
establishes a relation between displacement field $\textbf{u}$, pressure
$p$, body force $\textbf{F}=\rho \textbf{g}$ and boundary conditions $\textbf{bc}$, such
that we can consider $\textbf{u}$ to be uniquely determined from the others:
\begin{equation}
  \textbf{u} = \textbf{u}(p, \textbf{F}, \textbf{bc})
\end{equation}
Moreover, as \eqref{eq:forcebalance} is linear, the superposition principle applies.  As
such, if $p^0$ represents initial pressure and $p = p^0+\tilde p$ is a different
pressure, we have:
\begin{equation}
  \textbf{u}(p, \textbf{F}, \textbf{bc}) = \textbf{u}(p^0, \textbf{F}, \textbf{bc}) +
                                           \textbf{u}(\tilde p, 0, 0) 
\end{equation}
where $\textbf{u}(p^0, \textbf{F}, \textbf{bc})$ represents initial
deformation.  Assuming body forces and boundary conditions remain constant, we
can directly relate a change in displacements ${\boldsymbol{\tilde u} = \textbf{u}(p, \textbf{F},
\textbf{bc}) - \textbf{u}(p^0, \textbf{F}, \textbf{bc})}$ to
a change in pressure $\tilde p$:
\begin{equation}
  \boldsymbol{\tilde u} = \textbf{u}(\tilde p, 0, 0)
\end{equation}

When solving a discretized system based on \eqref{eq:forcebalance}, the pressure
field is defined using a linear combination of a set of basis functions 
$\{\phi_1,
...\phi_M\}$ and can be written: $p(\textbf{x}) = \sum_{i=1}^M
p_i\phi_i(\textbf{x})$.  Using the notation $\tilde p_i = p_i - p_i^0$, the superposition
principle allows us to write:
\begin{equation}
  \boldsymbol{\tilde u} = \sum_{i=1}^M\tilde p_i\boldsymbol{u}(\phi_i, 0, 0)
\end{equation}
The corresponding change in volumetric strain
$\tilde\epsilon=\nabla\cdot\boldsymbol{\tilde u}$ can thus be expressed:
\begin{equation}\label{eq:impulse-response}
  \tilde\epsilon = \sum_{i=1}^M\tilde p_i\Psi_i
\end{equation}
where $\Psi_i = \nabla\cdot\boldsymbol{u}(\phi_i, 0,0)$ represents the system's
volumetric strain response to pressure perturbation $\phi_i$.  This means that
knowledge of the set of (scalar) functions $\{\Psi_1, ...\Psi_M\}$ enables us to
insert \eqref{eq:impulse-response} into \eqref{eq:sys3} to obtain an equation
that only depends on pressure.  This equation can then be solved decoupled from
\eqref{eq:sys1} and \eqref{eq:sys2}, while still providing the same pressure
solution as the fully-coupled equation system.

In the context of volumetric strain, we will refer to the pressure
basis functions $\{\phi_i\}_{i=1...M}$ as \emph{impulse functions} and
$\{\Psi_i\}_{i=1...M}$ as \emph{response functions}.  Our proposed approach
consists of precomputing and storing a set of approximated volumetric strain response
functions that corresponds to our set of pressure basis functions.  At
simulation time, we can then solve \eqref{eq:sys3} using
\eqref{eq:impulse-response} to eliminate volumetric strain from the equation.

For the purpose of solving the pressure equation \eqref{eq:sys3} of our model
problem, we only need the values of the response functions
$\{\Psi_i\}_{i=1...m}$ restricted to $\Omega_{aq}$.  As the number of aquifer
grid cells in $\Omega_{aq}$ is generally much lower than the total number of
cells in $\Omega$, this allows us to cut down significantly the amount of
information that must be precomputed and stored.  A natural choice for the set
of impulse functions $\{\phi_i\}$ is to let $\phi_i$ represent a unit pressure
increase limited to aquifer grid cell $i$.  However, for the examples given in
the following section, a coarser set of impulse functions has been chosen, where
$\phi_i$ represents an unit pressure increase in an entire vertical column of
aquifer cells.  This particular choice relies on the assumption that
overpressure does not vary in a significant way across the vertical
thickness of the aquifer.  Given the large horizontal-to-vertical aspect ratio of a
typical aquifer, we find that this approximation tends to work well in practice,
while cutting back on the required amount of precomputation work.

In principle, the support of each response function $\Psi_i$ covers the entire
$\Omega_{aq}$.  However, for the physical conditions relevant to injection
scenarios, in particular the presence of a free-moving top (land or sea)
surface, $\Psi_i$ decays relatively quickly away from the support of the
corresponding impulse $\phi_i$.  This means it can be truncated at some distance
beyond which its value falls below some defined threshold.  The remaining
function $\tilde\Psi_i$ has local support, and can be rescaled so that
$\int_{\Omega_{aq}}\tilde\Psi_idx = \int_{\Omega_{aq}}\Psi_idx$.  It is worth
noting that if $\phi_i$ represents a unit pressure increase in cell $i$ and
$\Psi_i$ is discretized as a cell-wise constant function, then necessarily
$\int_{\Omega_{aq}}\Psi_idx = \bar c_{m,i}$, where $\bar c_{m,i}$ is the value
of $\bar c_m$ for cell $i$, as explained in the Appendix.  As such, the
truncated form $\tilde\Psi_i$ represents some intermediary between the fully
cell-restricted coefficient $\bar c_m$ and the function with full global support
$\Psi_i$, where all three entities represent the same total amount of aquifer
volumetric strain.

As a consequence, the use of $\bar c_m$ in combination with the previously
described local model can be understood as a limit case of using precomputed
responses with a sufficiently high threshold, so that the full support of
$\tilde\Psi_i$ falls within a single cell.  However, in this case we know that
the value of $\bar c_m$ for each and every aquifer cell can be determined by
solving equations \eqref{eq:sys1} and \eqref{eq:sys2} \emph{once}, as previously
explained.  Therefore, the required amount of precomputation is much less for
this limit case.

\begin{figure}[!h]
  \centering
  \includegraphics[width=.9\columnwidth, clip=true, trim = 50 0 0 0]{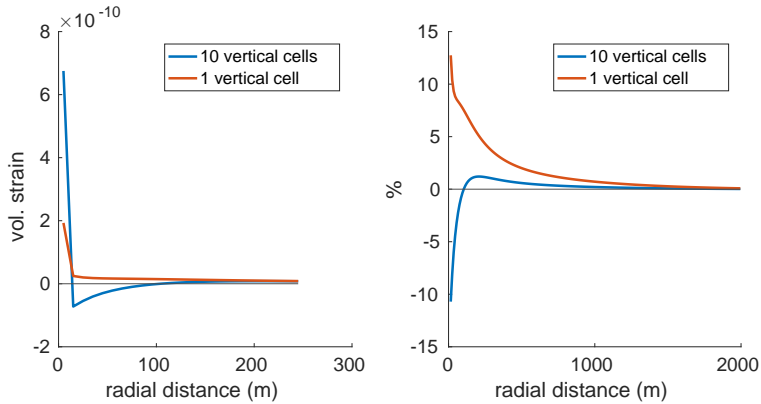}
  \caption{An example of a response function in 2D.  Volumetric strain plotted
    as a function of distance from a central pressure perturbation. The left
    plot includes the value of perturbed grid cell itself.  On the right plot,
    we have suppressed the perturbed grid cell and rescaled the curves to show
    percent-wise response magnitude compared with the central value.  A vertical
    aquifer resolution of 10 cells was used to compute the blue curves, whereas
    the red curves were computed using a vertical aquifer resolution consisting
    of a single cell.  With just one vertical cell, the bulging effect
    (c.f. Figure~\ref{fig:lifting-bulging}) is not properly
    captured.}\label{fig:impulses}
\end{figure}

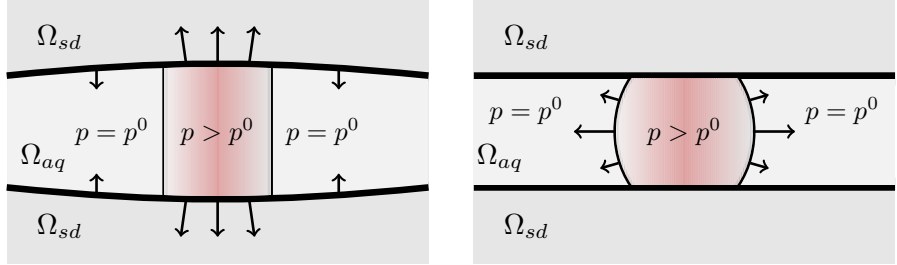
\begin{figure}
  \centering
  \begin{tikzpicture}[scale=0.7]
    
    \fill[gray!20!white] (-4,-1.5) rectangle (4,3.5);
    
    \draw[line width=5pt] (-4, 0) .. controls (-1, -0.3) and (1, -0.3) .. (4, 0)
    (4, 2) .. controls (1, 2.3) and (-1, 2.3) ..  (-4, 2);
    
    \fill[gray!10!white] (-4, 0) .. controls (-1, -0.3) and (1, -0.3) .. (4, 0) --
    (4, 2) .. controls (1, 2.3) and (-1, 2.3) ..  (-4, 2);
    
    \begin{scope}
      \clip (-4, 0) .. controls (-1, -0.3) and (1, -0.3) .. (4, 0) --
      (4, 2) .. controls (1, 2.3) and (-1, 2.3) ..  (-4, 2) -- cycle;
      \draw[line width=2pt] (-1, -0.3) rectangle (1, 2.3);
      \shade[left color=gray!10!white, right color=red!50!gray!50!white  ] (-1,-0.3) rectangle (0.01, 2.3);
      \shade[left color=red!50!gray!50!white  , right color=gray!20!white] (0,-0.3)  rectangle (1, 2.3);
    \end{scope}
    
    \draw(0,1) node{$p > p^0$};
    \draw(-2,1) node{$p=p^0$};
    \draw(2,1) node{$p=p^0$};
    \draw(-3.3, 0.5) node[scale=1.1]{$\Omega_{aq}$};
    \draw(-3, 2.8) node[scale=1.1]{$\Omega_{sd}$};
    \draw(-3, -0.8) node[scale=1.1]{$\Omega_{sd}$};
    
    \begin{scope}[->, line width=1pt]
      \draw (0,2.3) -- (0, 3);
      \draw (-0.6,2.3) -- (-0.7, 3);
      \draw (0.6, 2.3) -- (0.7, 3);
      \draw (0,-.3) -- (0, -1);
      \draw (-0.6,-.3) -- (-0.7, -1);
      \draw (0.6, -.3) -- (0.7, -1);
      
      \draw (2.3, 2.2) -- (2.3, 1.8);
      \draw (-2.3, 2.2) -- (-2.3, 1.8);
      \draw (2.3, -.2) -- (2.3, .2);
      \draw (-2.3, -.2) -- (-2.3, .2);
      
    \end{scope}
    
  \end{tikzpicture}
  \;\;
  \begin{tikzpicture}[scale=0.7]
    
    \fill[gray!20!white] (-4,-1.5) rectangle (4,3.5);
    \draw[line width=5pt] (-4, 0) --(4, 0)
    (4, 2) -- (-4, 2);
    
    \fill[gray!10!
    white] (-4, 0) rectangle (4, 2);
    
    pressure cell
    \begin{scope}
      \clip (-4, 0) rectangle (4, 2);
      \draw[line width=2pt] (0,1) ellipse (1.3 and 1.6);
      \clip (0,1) ellipse (1.3 and 1.6);
      \shade[left color=gray!10!white, right color=red!50!gray!50!white  ] (-1.3,-0.3) rectangle (0.01, 2.3);
      \shade[left color=red!50!gray!50!white  , right color=gray!20!white] (0,-0.3)  rectangle (1.3, 2.3);
    \end{scope}

    \draw(0,1) node{$p > p^0$};
    \draw(-3,1.4) node{$p=p^0$};
    \draw(3,1.4) node{$p=p^0$};
    \draw(-3.5, 0.5) node[scale=1.1]{$\Omega_{aq}$};
    \draw(-3, 2.8) node[scale=1.1]{$\Omega_{sd}$};
    \draw(-3, -0.8) node[scale=1.1]{$\Omega_{sd}$};
    
    \begin{scope}[->, line width=1pt]
      \draw (1.3,1) -- (2.1, 1);
      \draw (1.25,1.6) -- (1.6, 1.7);
      \draw (1.25,0.4) -- (1.6, 0.3);
      \draw (-1.3,1) -- (-2.1, 1);
      \draw (-1.25,1.6) -- (-1.6, 1.7);
      \draw (-1.25,0.4) -- (-1.6, 0.3);
      
    \end{scope}
  \end{tikzpicture}
  \caption{\emph{Left:} arching effect - local pressure perturbation pushes
    caprock upwards, causing surrounding aquifer region to
    stretch. \emph{Right:} bulging effect - local pressure perturbation causes
    the region to expand into the surrounding aquifer, causing compression of
    the neighboring rock.}
  \label{fig:lifting-bulging}
\end{figure}

A sample response function computed for a 2D model is presented in
Figure~\ref{fig:impulses}.  We here plot the height-averaged volumetric strain
response as a function of lateral distance from a unit pressure perturbation (1
Pa) in a 100 meter thick, horizontal aquifer at a depth of 1000 meters.  The
aquifer consists of soft rock with a Young's modulus value of 1 GPa, embedded in
a stiffer rock with a Young's modulus value of 10 MPa.  A high resolution is
used (10 m size grid blocks) in order to produce visually smooth curves.
Comparing the red and blue curves, it is clear that the response function looks
quite different depending on whether a single or ten layers were used in the
vertical discretization of the aquifer region.  This is because the full
mechanical effect of the pressure perturbation cannot be properly captured
without multiple vertical cell layers.  Using multiple different sets of input
parameters, we observe that response functions generally consist of an inner
negative part followed by a local positive maximum, and finally an attenuation
region where the response decays towards zero.  These qualitative features can
be explained as resulting from the counteracting effects of \emph{arching} and
\emph{bulging}.  The arching effect is produced by the elevated pressure of the
impulse region pushing vertically against the over- and underburden regions,
forcing them apart and causing nearby aquifer regions to expand
(Figure~\ref{fig:lifting-bulging}, left).  This effect becomes more spatially
spread out for higher stiffness ratios between the surrounding domain and the
aquifer rock.  On the other hand, there is a counteracting pushing and bulging
effect where the elevated pressure of the impulse region causes it to expand
laterally into its neighborhood, thereby compressing the nearby aquifer region
(Figure~\ref{fig:lifting-bulging}, right). At sufficiently short distances, this
effect dominates over the arching effect, resulting in a region with negative
volumetric strain.  At longer distances, the arching effect prevails, resulting
in the local peak and attenuation regions of the response function curves.

\section{Results}

In this section, we present a couple of cases where we compare the solutions
obtained using the fully coupled model (\emph{full model}) with those obtained
using our proposed method using precomputed response functions (\emph{PR
  model}) and using the standard one-way coupled approach based on the use of
local pore volume compressibility coefficients (\emph{local model}).

For the computations, we use the simulation framework provided by Matlab
Reservoir Simulation Toolbox \cite{MRST:2015b}.  Fluid flow is modeled using a
first order finite-volume upstream-weighted two-point flux approximation
numerical scheme, whereas mechanics is modeled using a discretization based on
first-order virtual elements \cite{gain2014} supplemented with approximate
higher-order energies to avoid artifacts that have been shown to arise for high cell
aspect ratios.  The interested reader is referred to \cite{Raynaud16:ecmor} or
\cite{Klemetsdal:2016} for related discussion.

In the full model, flow and mechanics equations are solved simultaneously as a
full linear system.  For the local model, the computed values of $\bar c_m$ are
used for pore volume compressibility coefficients.  Although the code is not
optimized for speed, the use of preconditioned iterative linear solvers (the
conjugated gradient method with incomplete Cholesky factorization; the
biconjugate gradients stabilized method with ILU factorization using threshold
and pivoting) significantly helped improve performance for all three models.

\subsection{Simple 2D example}

We begin by studying a very simple 2D injection example where fluid is injected
into a 100 m thick aquifer at a depth of 1000 m.  We consider pressure
development for a constant bottom hole overpressure of 1 MPa over a period of 50
days, and compare the results calculated from our three models (full model,
local model and PR model).  We use constant stress for lateral boundary
conditions.  Other relevant simulation parameters are listed in
Table~\ref{table:example1-params}.  This particular example was chosen to
illustrate the existence of two-way coupling effects in the near-well region.

\begin{table}[!htb]
  \centering
  \caption{Parameter values for simple 2D example}
  \label{table:example1-params}
  \begin{tabular}{ll}
    \hline
    Lateral extent                                     & 5 kilometers    \\
    Lateral resolution                                 & 31 cells        \\
    Vertical extent: overburden, aquifer, underburden  & 1000 m, 100 m, 800 m \\
    Vertical resolution                                & 10 cells, 5 cells, 10 cells \\
    Young's modulus, $\Omega_{aq}, \Omega_{sd}$          & 1 GPa, 10 GPa \\
    Poisson ratio, $\Omega$                            & 0.3 \\  
    Permeability, $\Omega_{aq}$                         & 100 mD \\
    Porosity, $\Omega_{aq}$                             & 0.3 \\
    Fluid compressibility                              & $4\cdot 10^{-5} \text{bar}^{-1}$ \\
    Fluid viscosity                                    & $0.8$ cP \\
    Well bottom-hole overpressure                      & 1 MPa \\
    Duration of simulation                             & 50 days \\
    Timestep size                                      & 1 day \\
    \hline
  \end{tabular}
\end{table}

The cutoff threshold used to compute the truncated response functions
$\tilde\Psi_i$ of the PR model is $10^{-2}$ times the maximum value of $\Psi_i$.

\begin{figure}[!h]
  \centering 
  \includegraphics[width=1\columnwidth]{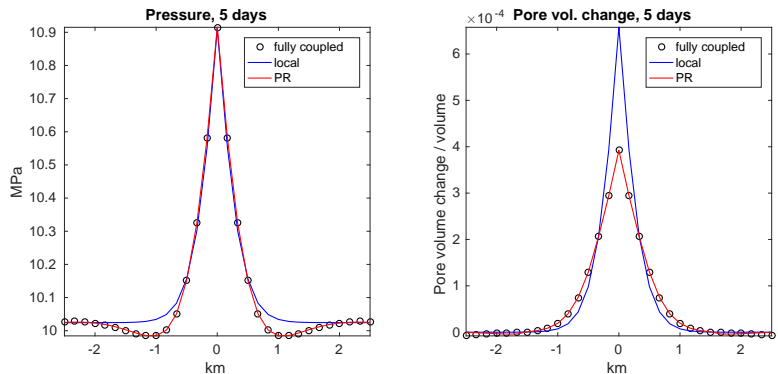}
  \caption{Pressure and pore volume change profiles at day 5 for the simple 2D
    injection example of this section}
  \label{fig:example1-year5}
\end{figure}

\begin{figure}[!h]
  \centering 
  \includegraphics[width=1\columnwidth]{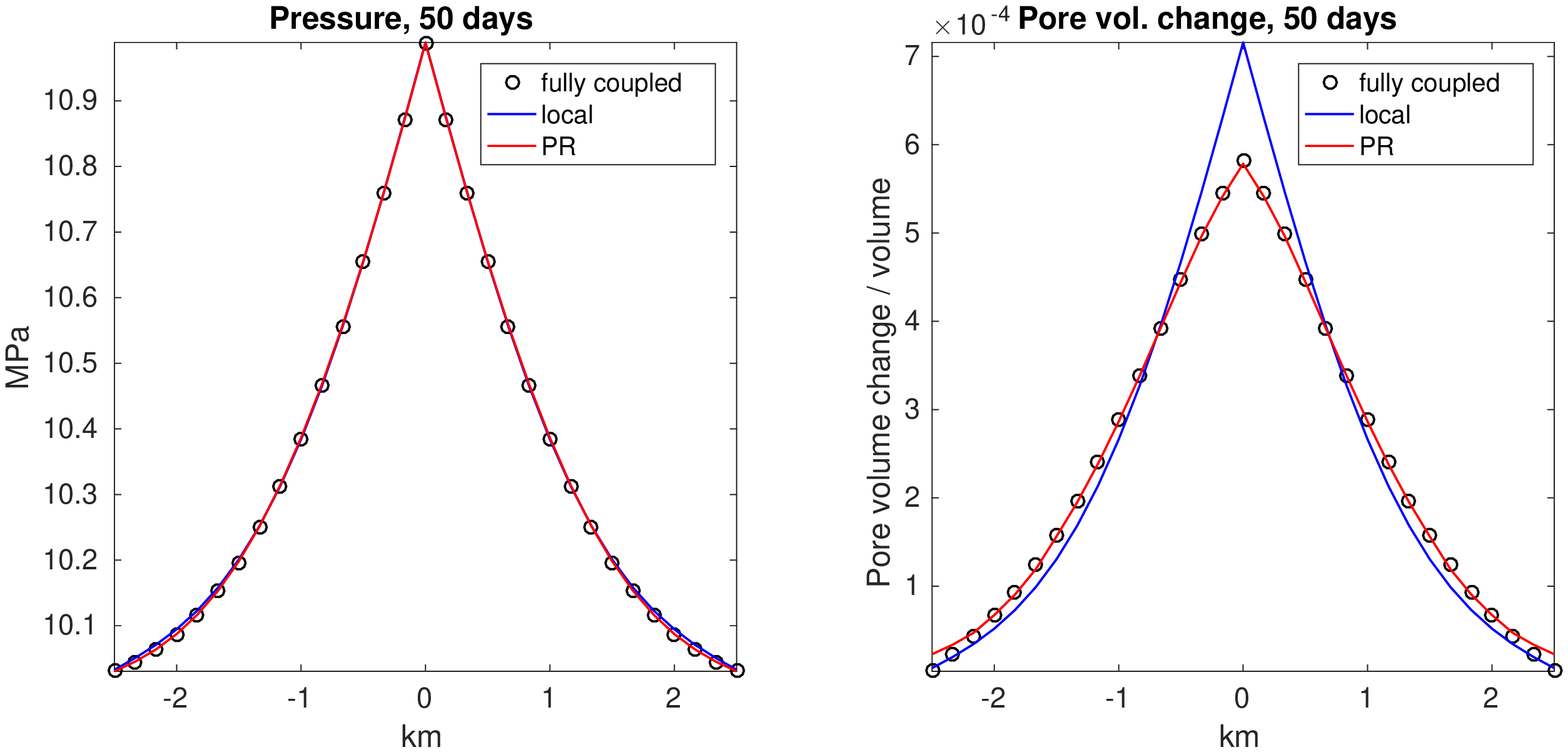}
  \caption{Pressure and pore volume change profiles at day 50 for the simple 2D
    injection example of this section}
  \label{fig:example1-year50}
\end{figure}

Aquifer pressure and corresponding pore volume changes for day 5 and day 50 are
plotted in Figure~\ref{fig:example1-year5} and \ref{fig:example1-year50}.  From
the left plot of Figure~\ref{fig:example1-year5}, we see that the PR model
reproduces quite well the result of the full model.  In particular, we observe
two zones with a local decrease in pressure, situated at either side of the
injection well, which are fully captured by the PR model but not by the local
model.  These pressure drops are associated with the arching effect where an
uplift of the caprock caused by high pressure around the well leads to
stretching and expansion of the aquifer rock in a wider area.  In our scenario,
the full model and PR model are thus able to predict a brief inflow and
accumulation of fluid into these regions at early times, an effect that cannot
be captured by the local model.

The right plot of Figure~\ref{fig:example1-year5} shows the fractional change in
aquifer pore volume (current minus pre-injection) for day 5, i.e. the
time-integrated first two terms of \eqref{eq:pressure-equation}.  For the full
model, the volumetric strain $\epsilon$ is immediately available since we have
access to the displacements at simulation time.  For the PR model, $\epsilon$ is
approximated using our precomputed response functions $\{\tilde\Psi_i\}_{i=1,
  ..., M}$ with
\eqref{eq:impulse-response}.  For the local model, we use \eqref{eq:uniaxial}.
In other words, we plot the pore volume changes as they are computed with
regards to the accumulation term of the pressure equation, not in terms of the
displacements that can be computed post-hoc from one-way coupling with mechanics system.

We note that the local model significantly overpredicts pore volume change
closest to the well, and slightly underpredicts it for about a kilometer before
reaching zero.  At very early times, the overprediction close to the well means
that the corresponding pressure is less than what is obtained with a two-way
coupled model.  When day 5 is reached, the increase in pore volume is still
significantly overestimated.  Nevertheless, the corresponding pressure is very
close to the correct value.  This is because the pore volume in itself does not
matter for the accumulation term in \eqref{eq:pressure-equation}, only its
change over time, which is already considerably smaller than for the first
couple of timesteps.

At 50 days (Figure~\ref{fig:example1-year50}), we see that the three models all
produce practically similar pressure profiles, whereas a large discrepancy
remains in terms of pore volumes.  It is interesting to note that the local
model, with its additional assumptions and computational simplicity, already
after 50 days produce results that are virtually identical to those given by the
two-way coupled models for this scenario.

Maximal and mean squared errors are presented in Table~\ref{table:example1-errors}.

\begin{table}
  \centering
  \caption{Maximal and mean squared errors for simple 2D example, in percent of
    total aquifer pressure variation (0.93 MPa for day 5 and 0.96 MPa for day 50) }
  \label{table:example1-errors}
  \begin{tabular}{|l|l|l|l|l|l}
    \multicolumn{1}{c}{} & \multicolumn{2}{c}{Day 5} & \multicolumn{2}{c}{Day 50}\\ \hline
    Model & Max error & Mean sq. error & Max error & Mean sq error\\ \hline
    Local & 5.32 & 3.19 & 0.79 & 0.43 \\
    PR    & 0.29 & 0.15 & 0.23 & 0.15 \\\hline
  \end{tabular}
  
\end{table}

\begin{figure}[!Ht]
  \centering
  \includegraphics[width=0.73\columnwidth]{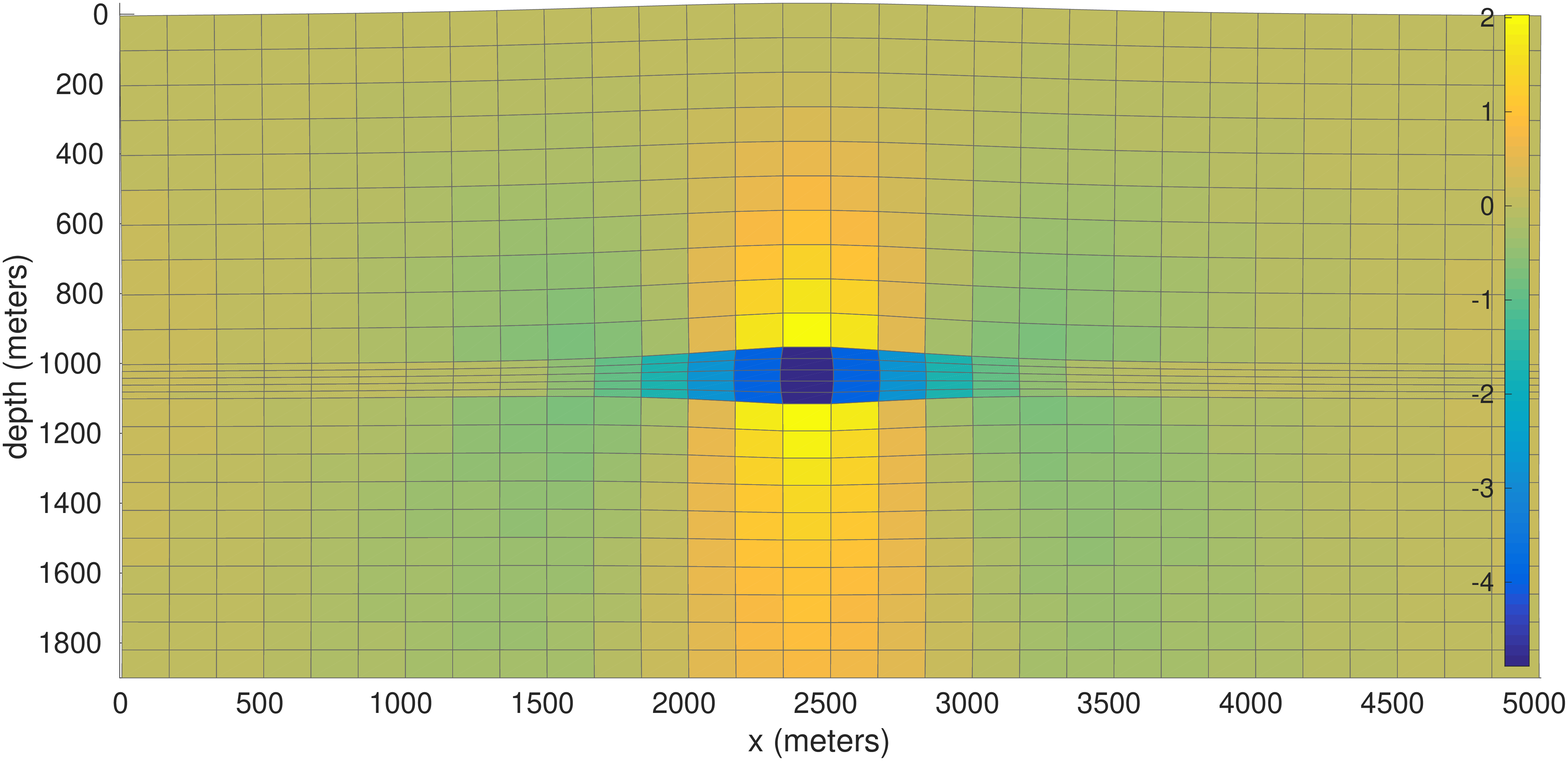}\\
  \vspace{2pt}
  \includegraphics[width=0.73\columnwidth]{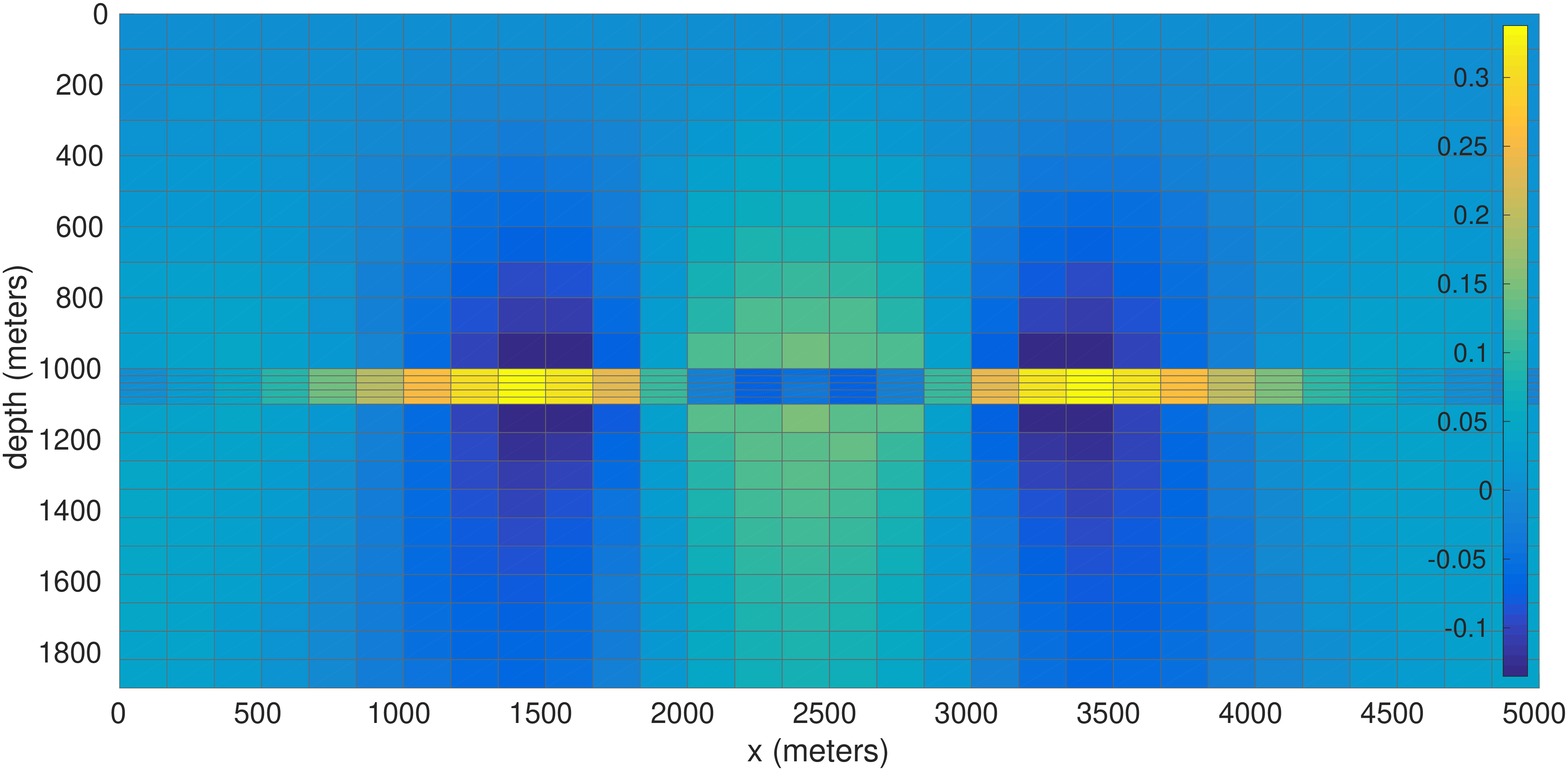}\\
  \vspace{2pt}
  \includegraphics[width=0.73\columnwidth]{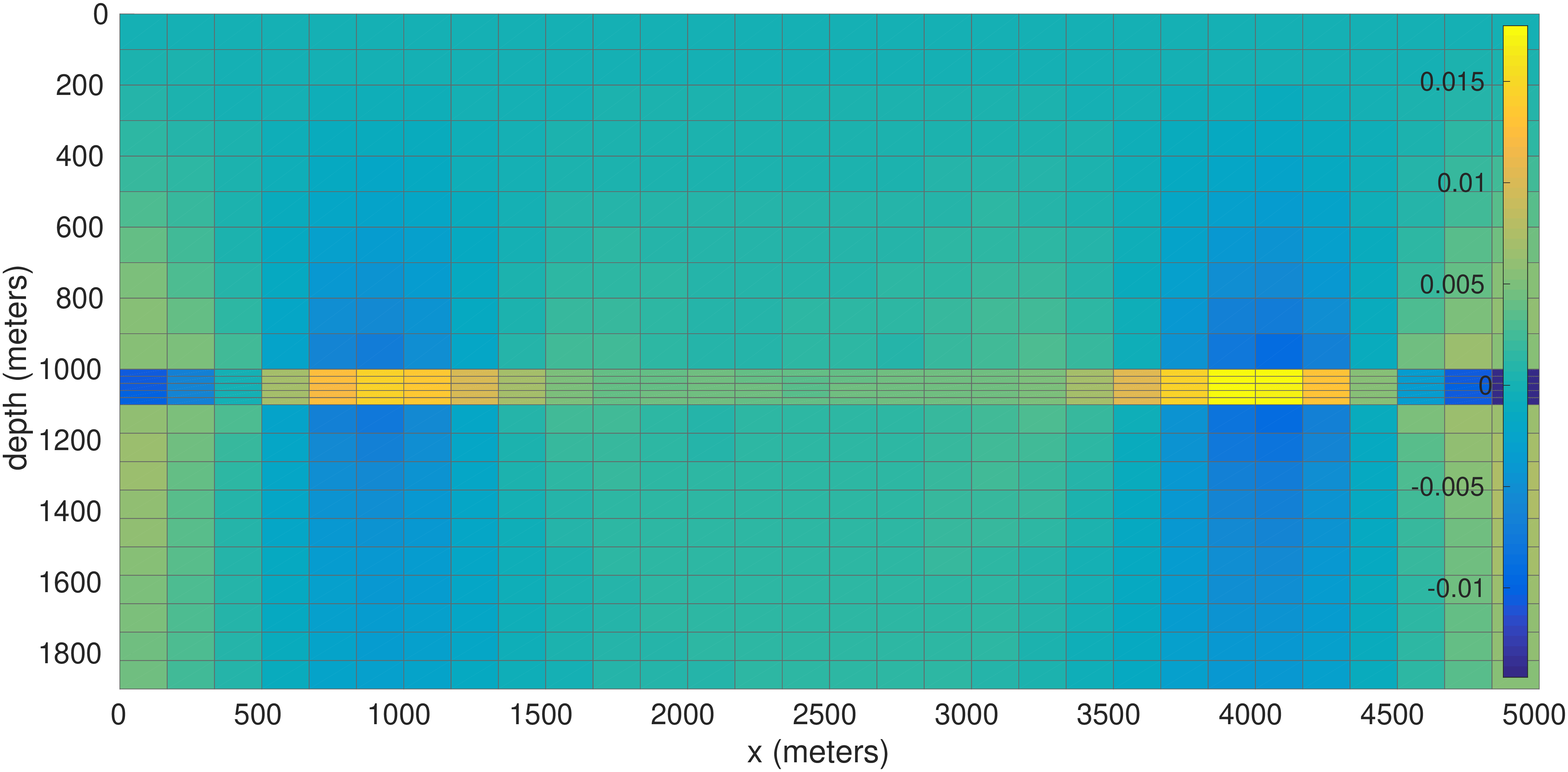}
  \caption{Change in vertical stress for aquifer, overburden and underburden at
    day 5 for the simple 2D injection example of this section.  Contrary to the
    sign convention used in the linear elasticity formulation in the previous
    section, we here employ the convention more prevalent in geomechanics
    literature where a positive sign indicates compressive stress. Units are in
    bar. \emph{Top}: Vertical stress change computed using the full model. Grid
    has been deformed to indicate the corresponding displacements (exaggerated).
    \emph{Middle}: Error between vertical stress computed from the full model
    and the local model. \emph{Bottom}: Error between vertical stress computed from
    the full model and the PR model.}
  \label{fig:example1-stress}
\end{figure}

Finally, we look at how discrepancies in pressure between the different models
impact the mechanics part of the solution.  As an example, we examine changes in
vertical stress.  The top plot in Figure~\ref{fig:example1-stress} visualizes
the change from initial state in vertical stress field at day 5 using the full
model.  The following two plots show the error between the vertical stress field
produced by the full model and those produced by the local and PR models,
respectively.  We note that the maximum error of the local model is about twenty
times higher than that of the PR model, for this timestep.  From the bottom
plot, we can see that the vertical stress field from the PR model closely
matches that of the full model in the middle two kilometers or so.  The largest
error is located at a distance of about 1500 meters, which is related to the
truncation of the response functions associated with cells closest to the
injection well.

\subsection{3D injection example}

We will now apply and compare our three methods on a 3D injection example that
borrows from the first benchmark problem in \cite{geocosa:2013}, whose details are
largely based on previous injection operations at the In Salah Carbon Capture
and Storage site \cite{rutqvist2010coupled}.  We chose this example because of
its relation to a real injection scenario where geomechanics issues have
played an important role.  However, for simplicity we restrict ourselves to
one-phase flow.
 
For the flow simulation, we consider injection into a uniformly 20 m thick,
horizontal aquifer. Our model covers 10 km of this aquifer in each lateral
direction ($\Omega_{aq}$).  The aquifer is located at a depth of 1800 m, with
fixed pressure imposed at lateral boundaries and an impermeable top and bottom.
In addition to the aquifer, the mechanical system includes the over- and
underburden.  The overburden consists of a ``shallow'' and a ``deep'' part with
different elastic properties.  The shallow part ($\Omega_{ob,s}$) extends from
the surface down to a depth of 900 meters, whereas the deep part
($\Omega_{ob,d}$) consists of the zone from 900 m to the aquifer depth of 1800
m.  The underburden ($\Omega_{ub}$) extends from the bottom aquifer boundary to
a depth of 4000 m, where we impose a a zero displacement boundary condition.
Lateral and top boundary conditions are of the fixed-stress type.  At the
beginning, we consider the aquifer and surrounding domain to be in mechanical
and hydrostatic equilibrium.  Further specifics on simulation grid and
parameters used are given in Table~\ref{table:example2-params}.

We simulate injection of fluid into the aquifer through a vertical well located
at the horizontal center of the modeled domain.  The well is perforated along
the full vertical extent of the aquifer.  In a first simulation case (CASE 1), we
consider a long-term, constant-rate injection for a total duration of 3 years (1
month timesteps).  The volumetric injection rate is set to 0.02 m$^3$/s at
aquifer conditions, which has been chosen to approximate the volumetric
injection rate of \co considered in \cite{rutqvist2010coupled}, taking the
inherent density differences between the injected fluids (water vs. \co) at
aquifer conditions into account.  In a second simulation case (CASE 2), we consider a
staggered injection pattern where 10 days of injection are followed by 10 days
of shut-off before the cycle is repeated, for a total simulated period of 90
days (5 day timesteps).

\begin{table}[!htb]
  \centering
  \caption{Parameter values for 3D injection example}
  \label{table:example2-params}
  \begin{tabular}{ll}
    \hline
    Lateral extent                                                                  & 10 x 10 kilometers    \\
    Lateral resolution                                                              & 21 x 21 cells         \\
    Layer thicknesses $\Omega_{ob,s}$, $\Omega_{ob,d}$, $\Omega_{aq}$, $\Omega_{ub}$   & 900
    m, 900 m, 20 m, 2180 m \\
    Vertical resolution $\Omega_{ob,s}$, $\Omega_{ob,d}$, $\Omega_{aq}$, $\Omega_{ub}$ & 3 cells, 4 cells, 3 cells, 7 cells \\
    Young's modulus $\Omega_{ob,s}$, $\Omega_{ob,d}$, $\Omega_{aq}$, $\Omega_{ub}$     & 1.5 GPa, 20 GPa, 6 GPa, 20 GPa \\
    Poisson ratio $\Omega_{ob,s}$, $\Omega_{ob,d}$, $\Omega_{aq}$, $\Omega_{ub}$       & 0.2, 0.15, 0.2, 0.15 \\  
    Aquifer porosity $\Omega_{ob,s}$, $\Omega_{ob,d}$, $\Omega_{aq}$, $\Omega_{ub}$    & 0.1, 0.01, 0.17, 0.15 \\
    Aquifer permeability                                                            & 13 mD \\
    Fluid compressibility                                                           & $4\cdot 10^{-5} \text{bar}^{-1}$ \\
    Fluid viscosity                                                                 & $0.32$ cP \\
    Fluid density                                                                   & 1000 kg/m$^3$\\
    Rock density                                                                    & 3000 kg/m$^3$\\
    Biot-Willis coefficient $\alpha$                                                & 1 \\
    \hline
  \end{tabular}
\end{table}

The cutoff threshold for truncating the response functions $\tilde\Psi_i$ of the
PR model was set to 10$^{-3}$ of the $\Psi_i$ maximum value.

\begin{figure}[H]
  \centering
  \includegraphics[width=1\columnwidth]{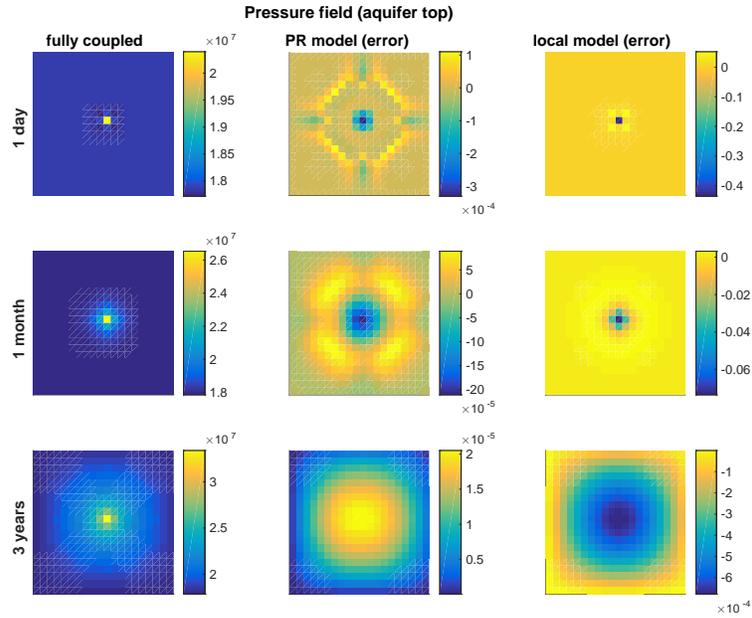}
  \caption{Top view of aquifer pressure at caprock level, for the continuous
    injection scenario (CASE 1).  Rows represent the status at one day, one
    month and three years after injection start, respectively.  The first column
    presents the pressure solution computed by the full (two-way coupled) model,
    with units in Pascal.  The second column shows the discrepancy between the
    full and the PR model, and the third column the discrepancy between the full
    and the local model, measured as fraction of total pressure change in
    aquifer (unitless).}
  \label{fig:case1-pressure}
\end{figure}

On Figure~\ref{fig:case1-pressure} we compare the outcomes of our three
different simulation approaches in terms of the computed aquifer pressure field
at selected points in time after injection start for CASE 1.  From the top row
we can see that after the first day, the local model underestimates the pressure
in the central grid cell by about 40\%, whereas the maximum error introduced by
the PR model is on the order of 10$^{-4}$, measured as a fraction of maximum
overpressure.  After 1 month, the error in the local model has shrunk to 6\%,
whereas the error using the PR model remains around 10$^{-4}$.  At 3 years (end
of simulation), the local model and PR model both produce results that are very
close to the fully coupled solution.

\begin{figure}[H]
  \centering
  \includegraphics[width=1\columnwidth]{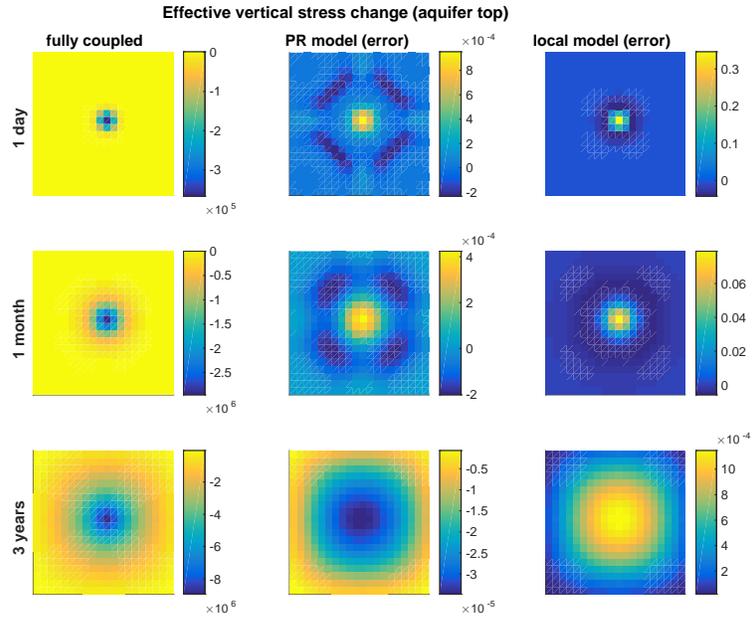}
  \caption{Top view of change in effective vertical stress in aquifer at caprock level,
    for the continuous injection scenario (CASE 1).  Rows represent the status
    at one day, one month and three years after injection start, respectively.
    On the figure, a positive sign indicates compressive stress.
    The first column presents the effective vertical stress computed by the full
  (two-way coupled) model, with units in Pascal.  The second column shows the
    discrepancy between the full and the PR model, and the third column the
    discrepancy between the full and the local model, measured as fraction of
    total vertical stress change in aquifer (unitless).}
  \label{fig:case1-stress}
\end{figure}

A similar trend can be seen in Figure~\ref{fig:case1-stress}, where we present
the corresponding changes in effective normal stress at the top of the aquifer
for CASE 1. Again, we see that for all timesteps, the error made by the PR model
remains on the order of 10$^{-4}$ or less of total stress change, whereas the
error resulting from the local model progressively shrinks from about 30\% after
day one to 6\% after a month and 10$^{-3}$ at the end of simulation.

\begin{figure}[H]
  \centering
  \includegraphics[width=0.49\columnwidth]{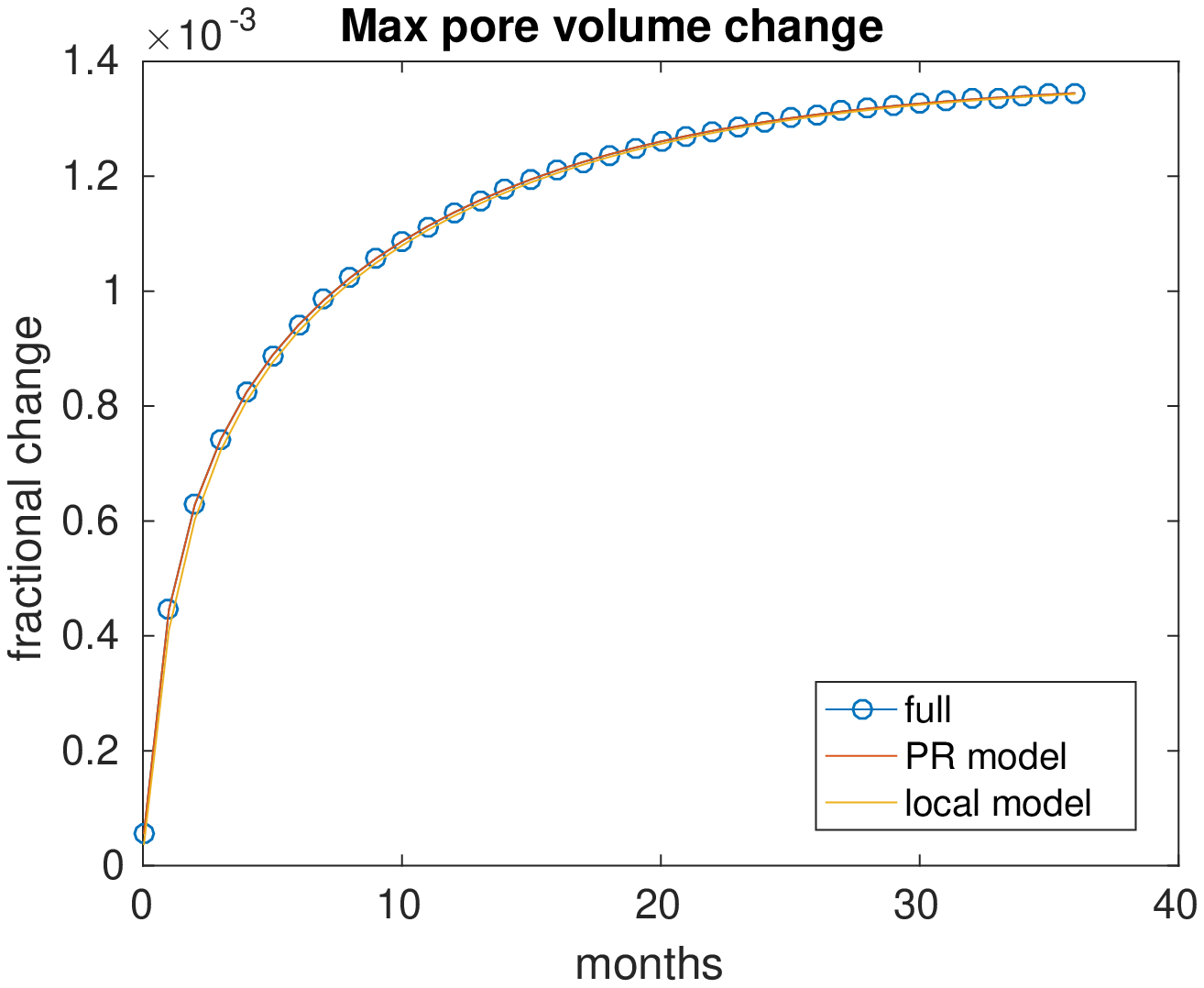}
  \includegraphics[width=0.49\columnwidth]{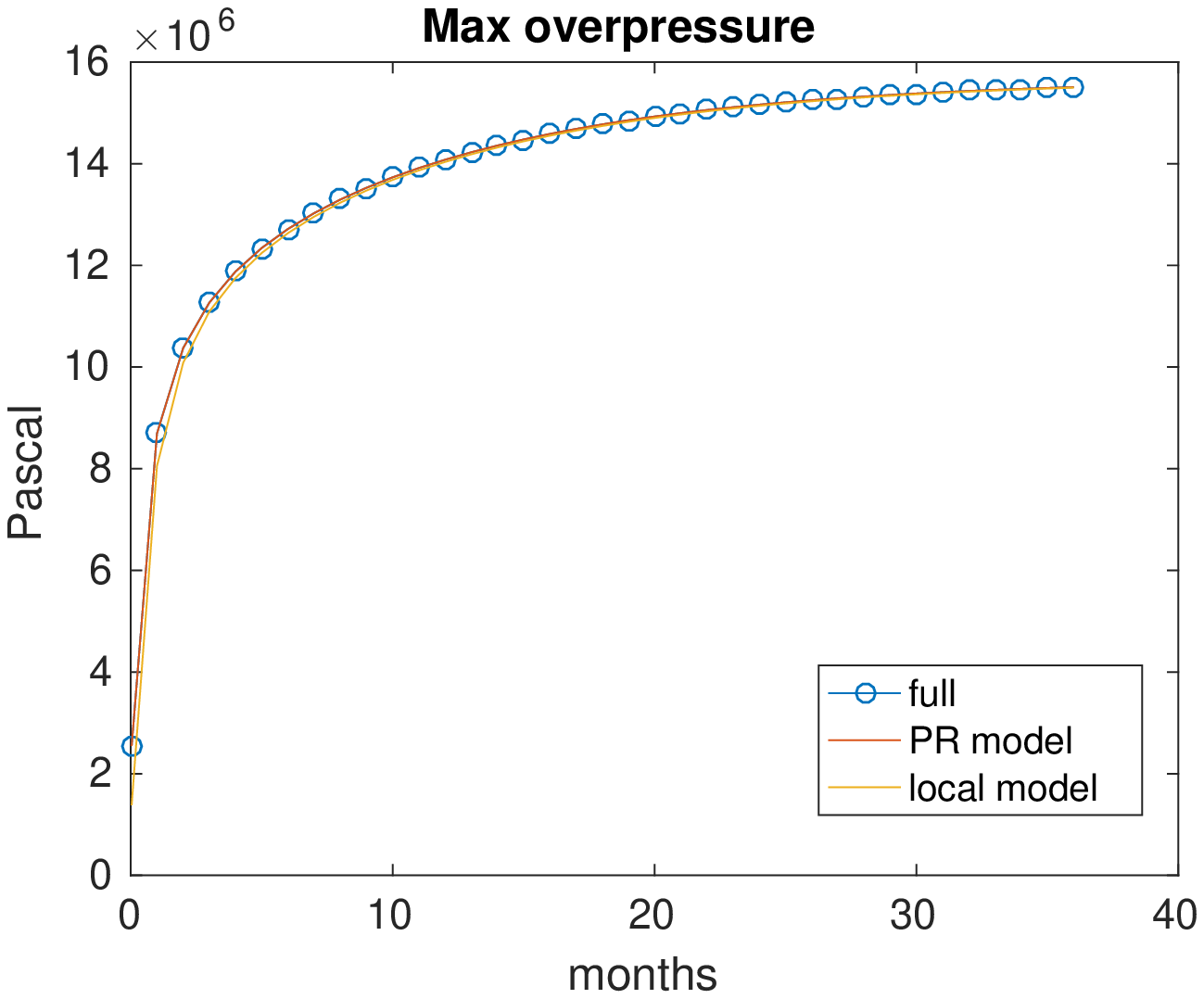}
  \caption{Temporal development of maximum pore volume change and maximum
    overpressure in aquifer for the continuous injection scenario (CASE 1).
    Maximum pore volume change expressed as fraction of bulk volume.}
  \label{fig:case1-temporal}
\end{figure}

On figure~\ref{fig:case1-temporal}, we track maximum change in pore volume and
maximum overpressure measured in the aquifer over time for CASE 1.  We see that
the three approaches produce curves that are very close, and only for the first
few timesteps is it possible to see any appreciable difference.  From all plots
presented so far for CASE 1, it seems clear that the PR model produce
consistent, good results, but that the local model \emph{also} produce good
results except for the very early period after injection start.

\begin{figure}[H]
  \centering
  \includegraphics[width=1\columnwidth]{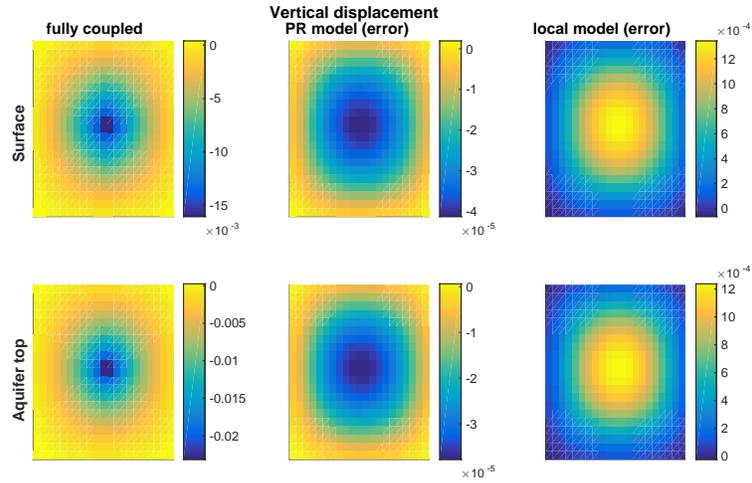}
  \caption{Vertical uplift at end of simulation period for the continuous
    injection scenario (CASE 1).  Upper row represents surface level and lower
    row aquifer caprock level.  The first column presents the vertical uplift
    computed by the full (two-way coupled) model, with units in meters.  The
    second column shows the discrepancy between the full and the PR model, and
    the third column the discrepancy between the full and the local model,
    measured as fraction of maximum uplift (unitless).}
    \label{fig:case1-uplift}
\end{figure} 

Vertical uplift around the injection area is a feature of the In Salah injection
operation that has been extensively studied and presented in past literature.  For our
CASE 1, the simulated uplift after 3 years is presented in
Figure~\ref{fig:case1-uplift}.  Regardless of approach, we arrive at a maximum
surface uplift of about 1.5 cm (z-axis oriented downwards).  This is consistent
with the range of values arrived at in \cite{rutqvist2010coupled} which is also
based on a coupled flow and geomechanics model.  However, it should be noted
that our example is limited to one-phase flow and thus not directly comparable.

\begin{figure}[H]
  \centering
  \includegraphics[width=1\columnwidth]{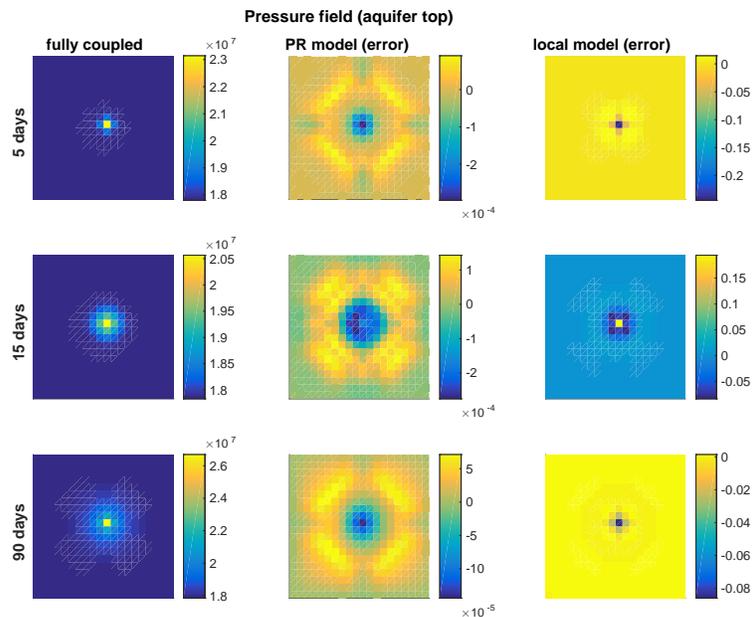}
  \caption{Top view of aquifer pressure at caprock level, for the staggered
    injection scenario (CASE 2).  Rows represent the status at one day, one
    month and three years after injection start, respectively.  The first column
    presents the pressure solution computed by the full (two-way coupled) model,
    with units in Pascal.  The second column shows the discrepancy between the
    full and the PR model, and the third column the discrepancy between the full
    and the local model, measured as fraction of total pressure change in
    aquifer (unitless).}
  \label{fig:case2-pressure}
\end{figure}

We now look at the corresponding results for CASE 2, which is a shorter-term
scenario with a non-uniform injection rate that prevents the system from
converging towards a long-term steady state. As such, we expect a priori that
the impact of using a fully-coupled model will be stronger.  The total period
modeled is 90 days.  Using a timestep duration of 5 days, the well switches
between on and off status each second timestep.

\begin{figure}[H]
  \centering
  \includegraphics[width=1\columnwidth]{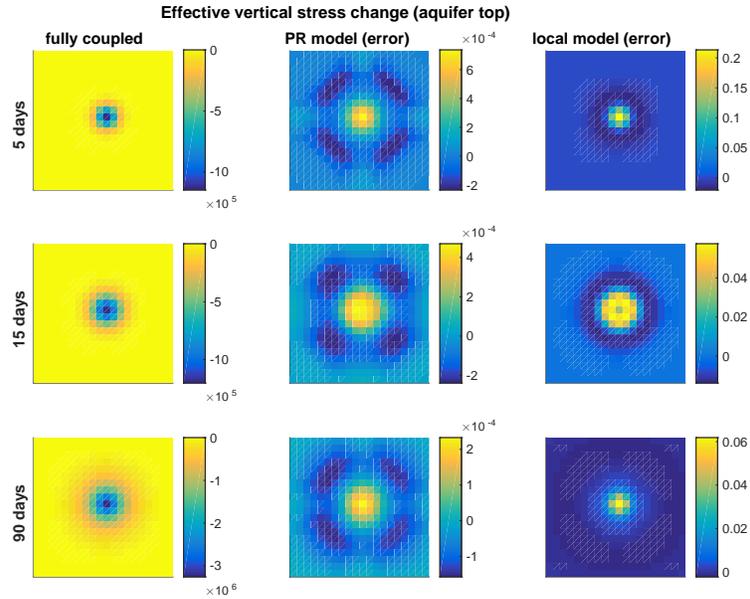}
  \caption{Top view of effective vertical stress in aquifer at caprock level,
    for the staggered injection scenario (CASE 2).  Rows represent the status
    at one day, one month and three years after injection start, respectively.
    On the figure, a positive sign indicates compressive stress.
    The first column presents the effective vertical stress computed by the full
  (two-way coupled) model, with units in Pascal.  The second column shows the
    discrepancy between the full and the PR model, and the third column the
    discrepancy between the full and the local model, measured as fraction of
    total vertical stress change in aquifer (unitless).}
  \label{fig:case2-stress}
\end{figure}

On Figure~\ref{fig:case2-pressure}, we plot aquifer pressure at 5 days, 15 days
and 90 days for CASE 2.  Similar to CASE 1, the discrepancy introduced by using
the PR model remains at the order of 10$^{-4}$ of maximum overpressure for all
three timesteps.  On the other hand, the local model error has a maximum error
of about 20\% after five days, and the error remains at 8\% at the end of
simulation.  The error is largely confined to the first few kilomters around the
well.  Similar observations regarding the overall error behavior from the PR and
local models can be made looking at the vertical stress change plots in
Figure~\ref{fig:case2-stress}, although the error in the local model is here
less strongly concentrated around the well.

\begin{figure}[H]
  \centering
  \includegraphics[width=0.49\columnwidth]{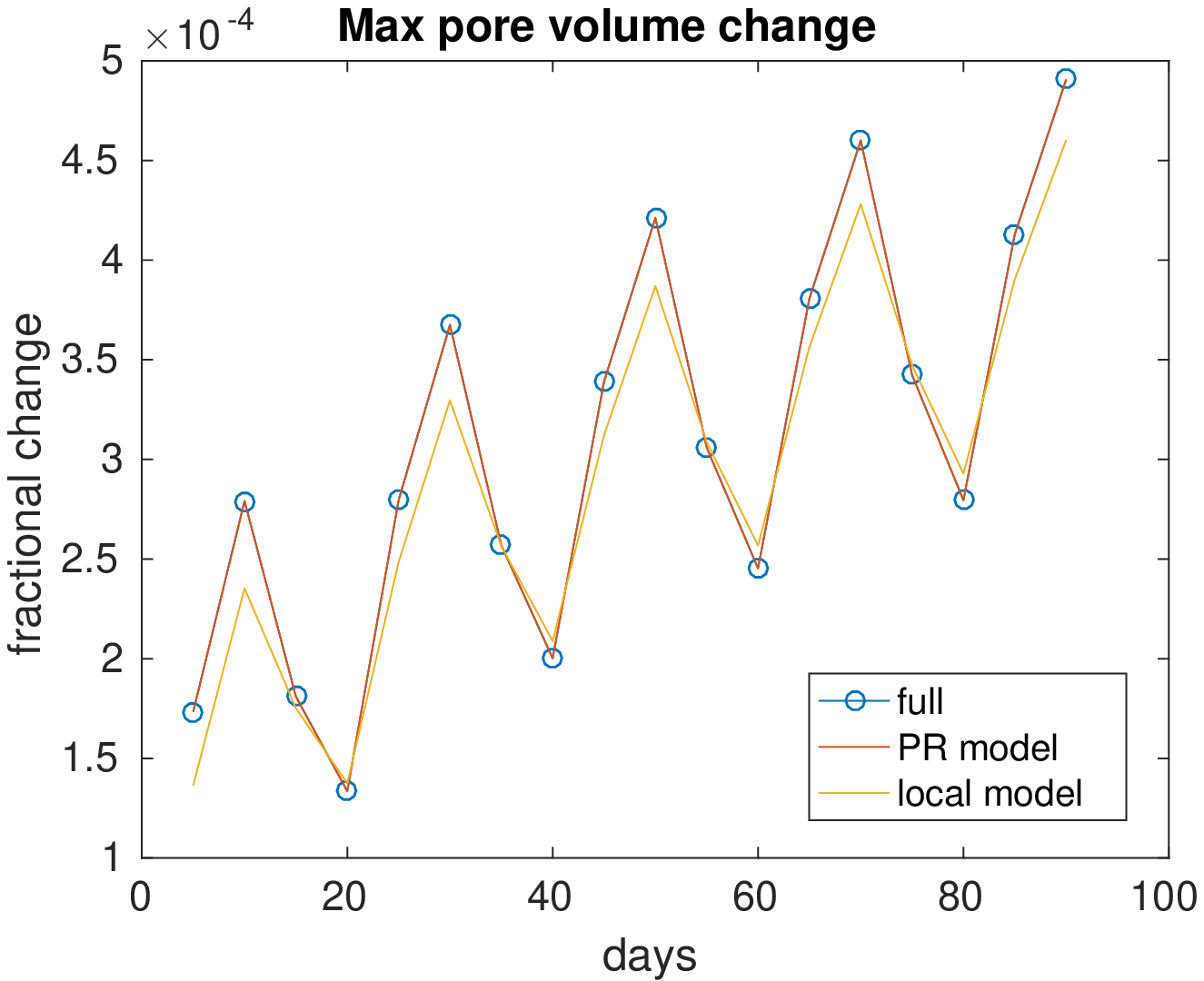}
  \includegraphics[width=0.49\columnwidth]{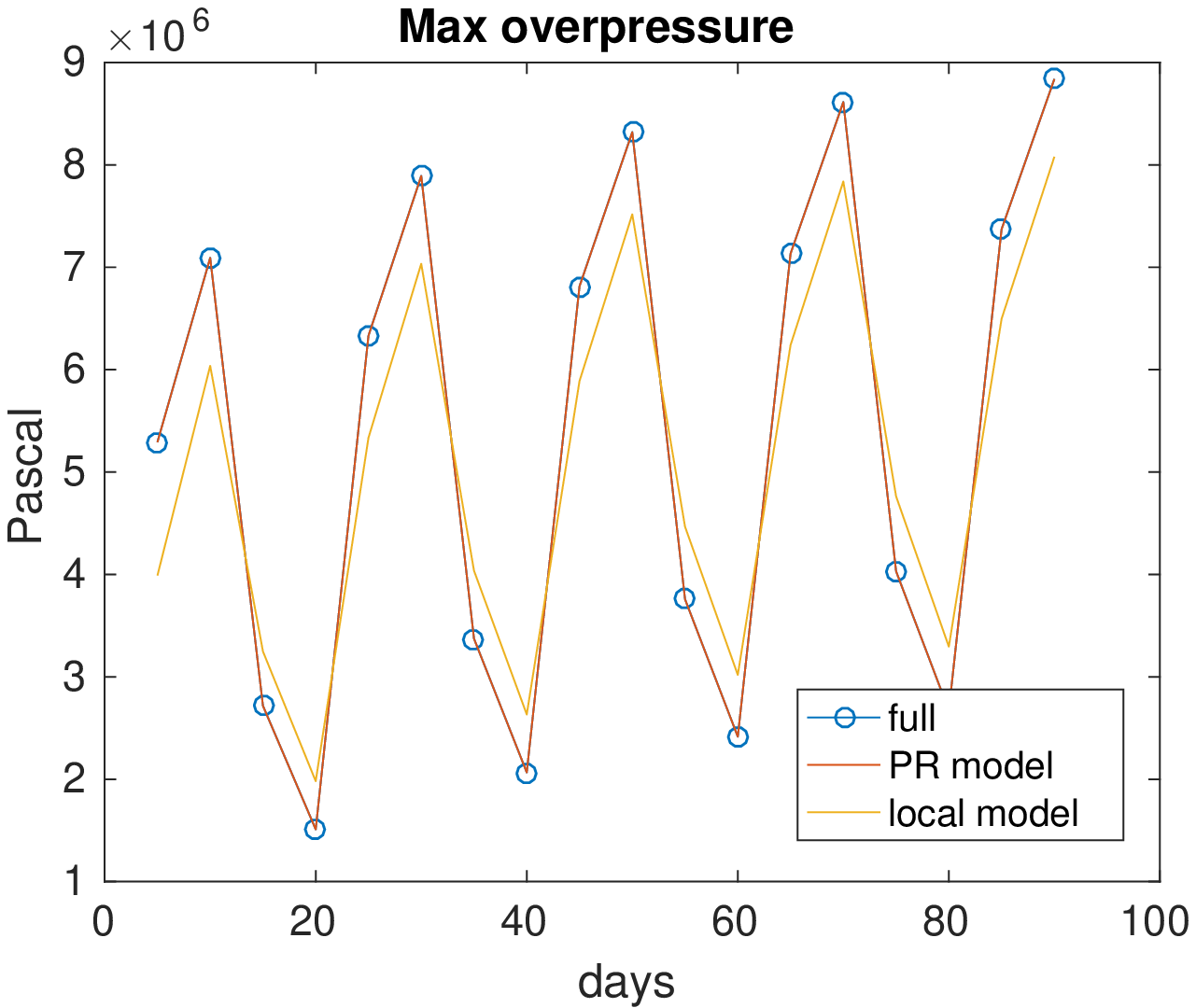}
  \caption{Temporal development of maximum pore volume change and maximum
    overpressure in aquifer for the staggered injection scenario (CASE 2).
    Maximum pore volume change expressed as fraction of bulk volume.}
  \label{fig:case2-temporal}
\end{figure}

\begin{figure}[H]
  \centering
  \includegraphics[width=0.49\columnwidth]{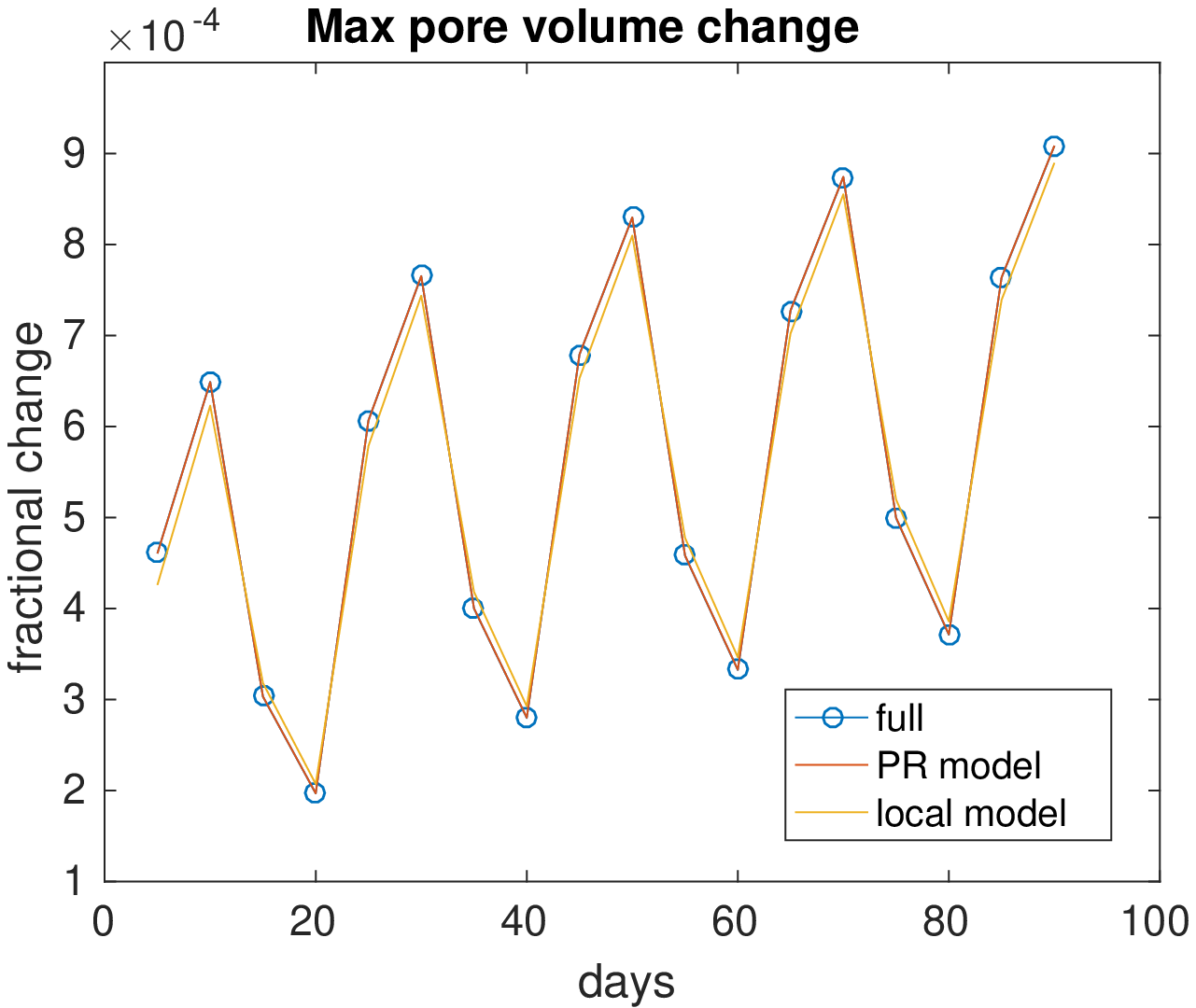}
  \includegraphics[width=0.49\columnwidth]{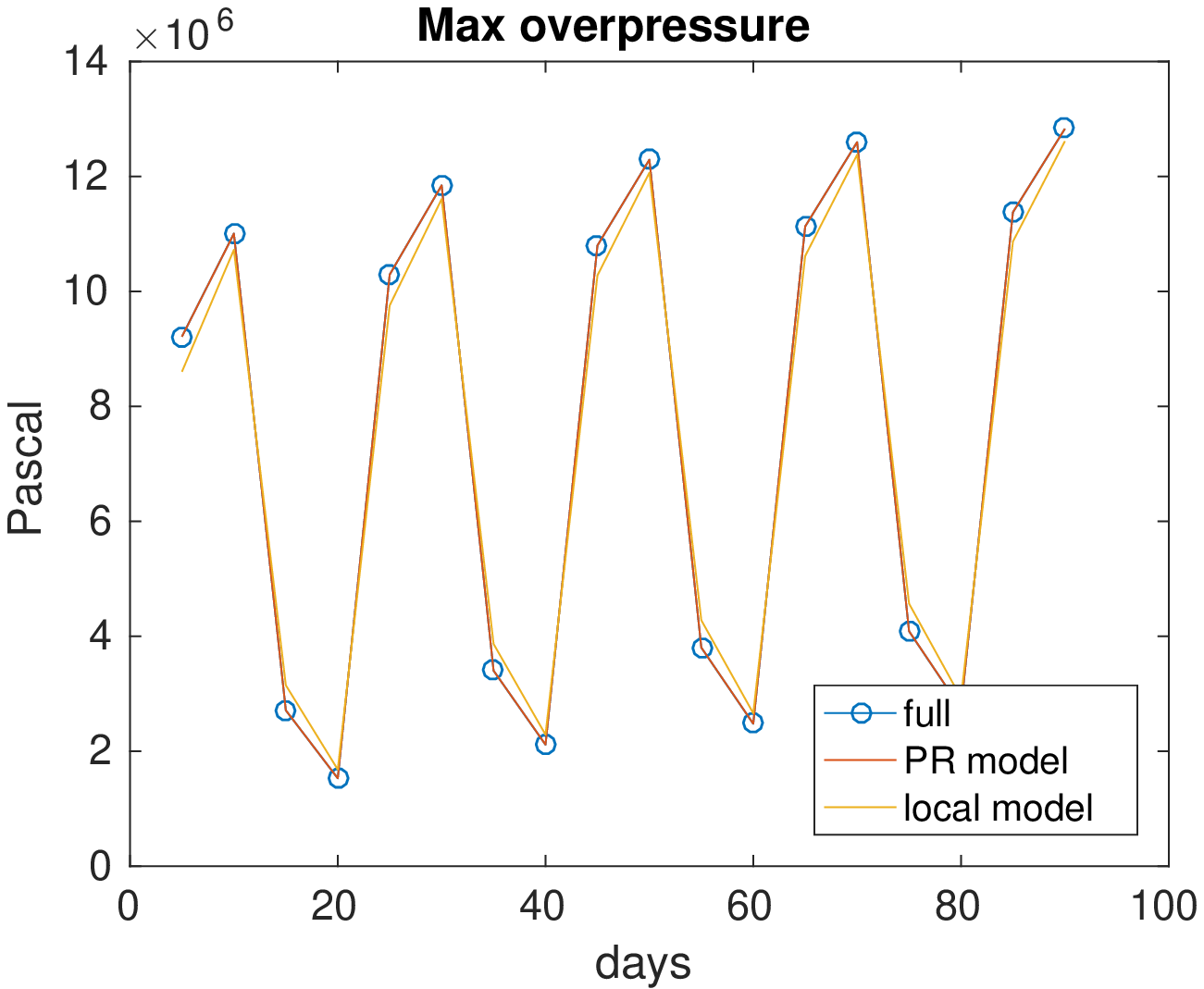}
  \caption{Temporal development of maximum pore volume change and maximum
    overpressure in aquifer for the staggered injection scenario (CASE 2), as
    simulated using a grid model with three times higher lateral resolution than
    in Figure~\ref{fig:case2-temporal}.  Maximum pore volume change expressed as
    fraction of bulk volume.}
  \label{fig:case2-temporal-hires}
\end{figure}

From the CASE 2 temporal plots of maximum pore volume change and overpressure in
Figure~\ref{fig:case2-temporal}, we note that the curves representing the
outcomes from the full and PR models remain close together, whereas the
curves representing the local model are noticeably different. Qualitatively, we
observe that the variations both in pore volume and pressure obtained from the
local model are weaker than those obtained from the other two models, which take
two-way coupling into account.  However, it should be noted that grid resolution
seems to have some impact on the magnitude of this discrepancy.
Figure~\ref{fig:case2-temporal-hires} present the corresponding curves after
re-running our simulations on a grid with about three times higher lateral
resolution (61 x 61).  As we can see, the difference between the local model and
the full and PR model is now considerably smaller.

\begin{table}[!htb]
  \centering
  \caption{Computational runtimes (seconds) for simulating CASE 1 and CASE 2
    using the full, PR and local model}
  \label{table:runtimes}
  \begin{tabular}{|l|c|r|r|r|r|}\hline
    Model & Case & Flow & Mechanics & Total & Timesteps \\ \hline
    Full  &    1 &    - &         - &   683.0 & 37 \\
    PR    &    1 &  8.8 &  85.0     &  93.8 & 37 \\
    Local &    1 &  4.4 &  84.6     &  89.0 & 37 \\  \hline
    Full  &    2 &    - &         - & 382.7 & 18 \\
    PR    &    2 &  6.0 &  38.7     &  44.7 & 18 \\
    Local &    2 &  2.7 &  46.7     &  49.4 & 18 \\  \hline
  \end{tabular}
\end{table}

The computational time for running CASE 1 and CASE 2 are presented in
Table~\ref{table:runtimes}.  The simulations were run on a standard workstation
equipped with a six-core Intel Core i7-3930K CPU and 20 GB RAM.  For the PR and
local models, runtime is broken down into ``Flow'' (time to run the flow
simulation) and ``Mechanics'' (time spent in post-hoc computation of
displacements, strains and stresses).  It is important to keep in mind the
prototype nature of the simulation software used.  Neither model has been
optimized with regards to speed. For instance, very conservative tolerances were
used for the fixed-stress split solver used to compute the fully coupled
solution, resulting in a number of linear and nonlinear iterations that might be
higher than necessary.  Nevertheless, the figures in the table suggest
that solving the system using the PR model represents an important gain in
efficiency compared with the full model, and remains computationally comparable
with the local model.  In our test cases, the computing time required for
solving the flow equation using the PR model was about twice the time spent
using the local model.  The difference is caused by a larger number of nonzeros
in the linear system to solve, as further discussed below.

Prior to running the simulations, the response functions had to be computed as a
preprocessing step.  This calculation took 147 seconds for our grid.  The
response functions can be reused for any simulation on this grid as long as
elastic moduli or mechanical boundary conditions do not change.

\subsection{Practical computational issues}

The precomputation of response functions, which happen once and can be
associated with the initial grid generation step, does incur
significant computational cost.  In theory, each response function $\Psi_i$
requires solving the linear system consisting of the discretized force balance
equations \eqref{eq:forcebalance} for a given right-hand side that represents the
impulse $\phi_i$.  For large aquifer models, the number of response functions
can be very high (one per aquifer cell or one per vertical column of aquifer
cells).  There are however some ways to mitigate this cost.

First, the linear system to solve, as represented by matrix $\bM$ in
\eqref{eq:discretized},  is symmetric, and can be efficiently
solved using a preconditioned iterative solver for symmetric systems.  Our
experience is that the conjugated gradient algorithm preconditioned with
incomplete Cholesky factorization works well in practice.  

Second, the task of computing response functions is trivial to parallelize.  All
computations work on the same immutable data and are completely independent
of each other.  The task of computing the full set of response functions can
therefore be spread out across as many computational cores as is available,
whether one is working on a standard modern multi-core laptop or using a high
performance computing cluster.  In contrast, the computational efforts required
for solving a fully-coupled system using iterative techniques cannot be
similarly parallelized.

Third, due to our use of thresholding, each truncated response function has
limited support.  Moreover, we know exactly what its rescaling factor is, namely
$\bar c_m$.  This allows us to compute several response functions at a time for
each single solution of the linear system, as long as they are sufficiently
spatially separated that their overlaps are practically negligible.  

All three strategies above were used in combination when computing the response
functions used in the numerical examples presented in this article.

It should be emphasized that thresholding has another important computational
aspect.  As the threshold is tightened, the support of each $\Psi_i$ widens,
leading to more coefficients to store but also a larger number of nonzeros in
the system matrix of the flow equation, i.e. the sparse approximation of the
matrix $\bE + \bS + \Delta t \bP$ in \eqref{eq:schur}.  As the number of
nonzeros of this matrix increases, iterative linear solvers will generally
require more computing time to solve the associated linear system.   The
threshold should therefore not be set too low.  In our experience, setting
threshold to 10$^{-3}$ appeared to be a good compromise for the examples
presented above.

\section{Conclusions}
Investigation of issues related to geological storage of \co will in many cases
require the ability to run a large number of numerical simulations within a
reasonable amount of time.  This necessitates the development of computationally
lightweight models.  The approach proposed in this paper, the use of precomputed
response functions, is an attempt to extend the current range of simplified
modeling capabilities for \co storage \cite{MRST:co2lab, NordbottenCelia}
to situations where full coupling between fluid flow and rock mechanics is
desired.
 
The work we have presented here suggests that this approach can reproduce, to a
good degree of approximation, the results from a fully-coupled fluid flow and
geomechanics simulation within the framework of linear poroelasticity, while
remaining comparable to a traditional flow simulation in terms of computational
requirements.  The work required for adapting traditional reservoir simulator
software to our approach amounts to simple modification of the accumulation
term in the mass conservation equation(s), so that the pore volume of a given
grid cell depends not only on pressure within that cell, but also on neighboring
pressures within a certain distance.  This can be seen as a generalization of
the common use of pore volume compressibility coefficients.

On the other hand, the approach where a traditional flow simulation is one-way
coupled with a mechanics solver seems to perform very well in many situations,
Moreover, the use of an numerically computed $\bar c_m$ does allow to account to
a certain extent for material heterogeneities and different boundary conditions,
and the corresponding local model can indeed be seen as a limit case of our
proposed approach based on precomputed responses.

In our experience the need for full coupling thus appears to be most relevant for 
models of limited spatial extent, short-term temporal scale, or to cases that
involve significant temporal variation.  As we saw for the 3D example above, the
use of varying injection rates leads to a system that does not rapidly approach
quasi-steady state conditions, and thus the non-local geomechanical impact on
flow persists over time.  Regarding this point, it is worth to mention that
real-life injection operations will amost always suffer from transiently varying
injection, whether it be for regular maintenance, or by design, as is the case
of offshore injection by ship.

It is also possible that more complex simulation models than those examined
herein will exhibit stronger coupling behavior, e.g. when using
strain-dependent permeability.  Whenever this coupling is strong enough to
justify the extra computational and algorithmic overhead, the use of precomputed
response functions can allow this coupling to be accounted for within the flow
simulator itself in a computationally efficient way.

In the context of \co storage, geomechanical studies are often associated with
the need to understand the risk posed by nearby faults, whose mechanical
behavior might not be properly described within a linear poroelastic framework.
In that case, a hybrid model can be envisaged where the aquifer and rock matrix
away from the fault is described using the linear theory, whereas the fault
itself is modeled using nonlinear constitutive relationships.  In such
situations, precomputed response functions might still be used to describe the
mechanical behavior of the part of the aquifer system that behaves linearly,
away from the fault itself.  However, this has not yet been tried and remains a
topic for future research.

We conclude that the use of precomputed response functions offers the
possibility to include the full impact of two-way geomechanical coupling into a
stand-alone flow simulation in a computationally efficient way.  The method
appears to work well and to offer significant advantages, in particular for
cases where many simulations need to be carried out.  Further demonstration of
the practical utility of the method will however require clear use cases that
are both relevant for \co storage and not already sufficiently well described by
the considerably simpler local approach based on pore volume compressibility
coefficients, in particular when combined with the use of  $\bar c_m$.

\section{Acknowledgments}
This work was funded by the MatMoRa-II project, Contract no. 21564, sponsored by
the Research Council of Norway and Statoil ASA.
\section{Appendix}
\subsection{Analytical solution of the force equilibrium equation on an
  unbounded domain}
This argument is adapted from the discussion at the beginning of Chapter 5 of
\cite{wang2000theory}.  We consider the poroelastic force equilibrium equation
on an unbounded 3D domain:
\begin{equation}\label{eq:forcebalance2}
   \nabla\cdot (G\nabla\textbf{u}) + \nabla\left((K +
   \frac{1}{3}G)\nabla\cdot\textbf{u}\right) = \alpha\nabla p - \textbf{F}\;\;\text{in}\;\; \Omega = \mathbb{R}^3
\end{equation}
where \textbf{F} represent some body force.  Using $c_m = \frac{\alpha}{K_v}$
along with standard relations between poroelastic parameters, it can be readily
verified that a particular solution to \eqref{eq:forcebalance2} is given by:
\begin{equation}
u_i = \frac{\partial\Phi}{\partial x_i}
\end{equation}
if $\Phi$ is some scalar potential that satisfies:
\begin{equation}
  \nabla^2\Phi = c_m p
\end{equation}
However, the Laplacian of $\Phi$ is then the volumetric strain, since:
\begin{equation}
  \nabla^2\Phi = \frac{\partial u_k}{\partial x_k} = \epsilon_{kk} = \epsilon
\end{equation}
Thus, the solution obtained implies that $\epsilon = c_m p$.  In other words,
volumetric strain at any given point is directly proportional to the pressure at
that point, and consequently the relation between volumetric strain and pressure is
purely local, assuming an unbounded domain and constant values of $K, G$ and $\alpha$.

\subsection{Demonstration that $\int_{\Omega_{aq}} \Psi_i dx = \bar c_m$ on a discrete grid} 

We depart from expression \eqref{eq:impulse-response}.  We assume that $\Psi_i$
represents the volumetric strain response to a unit pressure increase in cell
$i$ (i.e. the impulse function $\phi_i$ equals 1 on cell $i$ and zero
elsewhere).  We further consider that $\tilde\epsilon$ and $\Psi_i$ are
cell-wise constant, in which case we write $\tilde\epsilon_j$ and $\Psi_{i,j}$
to represent their values on grid cell $j$.  \eqref{eq:impulse-response} can
then be written:
\begin{equation}
  \tilde\epsilon_j = \sum_{i=1}^M\tilde p_i\Psi_{i,j}
\end{equation}

As we have previously specified $\bar c_{m,j}$ to represent the local volumetric strain
response in cell $j$ for a aquifer-wide unit pressure increase, we obtain:

\begin{equation}
  \bar c_{m, j} = \sum_{i=1}^M\Psi_{i,j}
\end{equation}

However, from its definition, $\Psi_{i,j}$ is also element $(i,j)$ of matrix $\bE$
in \eqref{eq:schur}, which is symmetric.  Hence:

\begin{equation*}
  \bar c_{m, j} = \sum_{i=1}^M\Psi_{i,j} = \sum_{i=1}^M\Psi_{j,i} = \int_{\Omega_{aq}}\Psi_j dx
\end{equation*}

In other words, the total volumetric strain across the aquifer resulting from a
pressure increase in cell $j$ \emph{also} equals $c_{m,j}$.  When $c_m$ is used as a pore
volume compressibility coefficient in a decoupled flow simulation, this is
equivalent to considering that all volumetric strain caused by a pressure
increase in a cell $i$ is concentrated to that cell.

We now present an informal argument to suggest a similar result in the
continuous case.   We start by considering \eqref{eq:forcebalance2} on the form:
\begin{equation}
  \mathcal{L}(\tilde\bu) = \alpha\nabla\tilde p
\end{equation}
where $\mathcal{L}$ is the linear, self-adjoint differential operator associated
with the left hand side of \eqref{eq:forcebalance2} with specified boundary
conditions, and we ignore the body force $\textbf{F}$ since we are only
concerned with changes in $\bu$ (which we denote $\tilde\bu$), with respect to changes in
$p$ (which we denote $\tilde p$).  If we assume that the zero-displacement (Dirichlet) part of the
boundary has nonzero measure, Korn's inequality assures a trivial kernel of
$\mathcal{L}$.  The associated Green's function $G(x, \xi)$ satisfies:
\begin{equation}
  \mathcal{L}G(x, \xi) = \delta(x-\xi)
\end{equation}
where $\delta$ denotes the delta function.  $G(x, \xi)$ is a rank two tensor
that expresses the displacement vector at $u$ caused by a force applied at
$\xi$.  Since $\mathcal{L}$ is self-adjoint, $G$ is symmetric, i.e. $G(x, \xi) =
G(\xi, x)$.  This also immediately follows from Maxwell's reciprocity relation.

The displacement $\tilde u$ caused by a pressure variation $\tilde p$ in the
domain can be written:
\begin{equation}\label{eq:green-displacement}
  \tilde u(x) = \int_\Omega G(x,\xi)\nabla \tilde p(\xi) d\xi =
  -\int_\Omega\nabla_\xi\cdot G(x,\xi)\tilde p(\xi)d\xi
\end{equation}
We have here loaded the differential operator $\nabla$ onto $G$ in the $\xi$
argument.  For a unit pressure increase across $\Omega$, we thus have:
\begin{equation}
  \tilde u(x) = -\int_\Omega\nabla_\xi\cdot G(x,\xi)d\xi
\end{equation}
Using the symmetry of $G$, the divergence of $\tilde u$ at $x$ for a unit
pressure increase across $\Omega$ can be developed as follows:
\begin{align*}\label{eq:green-div}
  \nabla\cdot\tilde u(x) &= \int_\Omega\nabla \cdot u(s)\delta(s-x) ds \\
              &= -\int_\Omega\nabla_s\cdot\left(\int_\Omega\nabla_\xi\cdot
              G(s,\xi)d\xi\right)\delta(s-x) ds\\
            &= -\int_\Omega\nabla_\xi\cdot\left(\int_\Omega\nabla_s\cdot G(\xi,
            s)\delta(s-x) ds\right) d\xi \numberthis
\end{align*}
Comparing with \eqref{eq:green-displacement}, the expression inside the
parentheses on the last line of \eqref{eq:green-div} expresses the displacement
at $\xi$ resulting from a delta pressure at $x$.  The full expression of the last
line thus represent the integrated divergence of displacement (volumetric
strain) across $\Omega$ associated with a delta pressure at $x$.

Thus the volumetric strain at $x$ resulting from unit pressure increase across $\Omega$
equals the integrated volumetric strain across $\Omega$ resulting from a
delta pressure in $x$.

%
%
\bibliography{bibfiles/co2,bibfiles/misc,diverse}

\end{document}